\documentclass[prd,12pt,tightenlines]{revtex4}

\usepackage{amsfonts,amssymb,amsthm}
\usepackage{color,epsfig}

\newcommand{\C}{{\mathbb C}}

\newcommand{\R}{{\mathbb R}}
\newcommand{\Z}{{\mathbb Z}}

\newcommand{\SU}{{\rm SU}}

\newcommand{\be}{\begin{equation}}
\newcommand{\ee}{\end{equation}}
\newcommand{\beq}{\begin{eqnarray}}
\newcommand{\eeq}{\end{eqnarray}}

\newcommand{\bra}{\langle}
\newcommand{\ket}{\rangle}

\newcommand{\bg}{{\bf g}}
\newcommand{\rd}{\mathrm{d}}

\def\lsim{\
  \lower-2.0pt\vbox{\hbox{\rlap{$<$}\lower5.5pt\vbox{\hbox{$\sim$}}}}\ }
\def\gsim{\
  \lower-2.0pt\vbox{\hbox{\rlap{$>$}\lower5.5pt\vbox{\hbox{$\sim$}}}}\ }

\newtheorem{prop}{Proposition}[section]

\newtheorem{lemma}[prop]{Lemma}

\begin{document}

\title{ A new Spin Foam Model for 4d Gravity}

%\date{July 2007}

\author{Laurent Freidel\,$^a$ and Kirill Krasnov\,$^{a,b}$ }
\affiliation{$^a$ Perimeter Institute for Theoretical Physics,
Waterloo, N2L 2Y5, Canada. \\ $^b$ School of Mathematical
Sciences, University of Nottingham, Nottingham, NG7 2RD, UK}

\begin{abstract}
Starting from Plebanski formulation of gravity as a constrained BF theory we propose a 
new spin foam model for 4d Riemannian quantum gravity that generalises 
the well-known Barrett-Crane model and resolves the inherent to it ultra-locality problem. 
The BF formulation of 4d gravity possesses two sectors: gravitational and topological ones. 
The model presented here is shown to give a quantization of the gravitational sector, and 
is dual to the recently proposed spin foam model of Engle et al. which, we show, 
corresponds to the topological sector. Our methods allow us to introduce the Immirzi 
parameter into the framework of spin foam quantisation. We generalize some of our 
considerations to the Lorentzian setting and obtain a new spin foam model in that context as well.
\end{abstract}

\maketitle

\section{Introduction}

The spin foam approach to quantum gravity was born about a decade
ago as a result of cross-fertilization between several different
fields: loop quantum gravity, topological quantum field theory and
simplicial gravity. On one hand, this approach is supposed to give
a spacetime covariant version of the picture of quantum geometry
arising in loop quantum gravity. On the other hand, it attempts to
apply methods and ideas that are known to be very powerful in the
context of 3-dimensional quantum gravity to the problem of
quantization of general relativity (GR) in 4 spacetime dimensions.
At the same time, the models considered are in the
simplicial setting, and are either derived from or inspired by a
discretization of some classical gravity action.

The most well-known model --that due to Ponzano and Regge \cite{PR}-- 
appeared in the sixties in the context of three-dimensional quantum gravity, 
thirty years before the term ``spin foam models'' and twenty years before 
the first works on loop quantum gravity. 
In the context of four spacetime dimensions the notion of (what we now call)
a spin foam as a history of spin network evolution was introduced,
from the perspective of loop quantum gravity, in
\cite{Reisenberger:1994aw,Reisenberger:1996pu}. A new era for
the subject was opened by an important work 
by Barbieri \cite{Barbieri:1997ks}, who realized that quantum states of loop quantum 
gravity can be understood (and heuristically derived) by applying the quantization 
procedure to the usual geometric tetrahedron in 3 spatial dimensions.
Even though this work may have not
received a proper credit in the literature, it was extremely influential in that
it suggested that at least to some extent quantum gravity may be about quantizing 
geometric structures. This idea was then quickly applied to the problem of
quantization of a geometric simplex in $\R^4$, with a breakthrough result being the construction
of the now famous Barrett-Crane model \cite{Barrett:1997gw}. 
A relation between this model and various versions of the simplicial gravity was 
explored in works \cite{Reisenberger:1998fk}, \cite{De Pietri:1998mb} and \cite{Freidel:1998pt}
showing that the Barret-Crane model is related to a
formulation of gravity as a constrained BF theory.

The key notion arising in the Barrett-Crane model is that of
simple (or class I) representations of the Lorentz group in four
dimensions. In work by Freidel, Krasnov and Puzio \cite{Freidel:1999rr}
these representations were understood as coming from the harmonic
analysis on the certain homogeneous group space, a crucial fact
that opened a way to the Lorentzian generalization \cite{Barrett:1999qw}
of the original Barrett-Crane model in Euclidean signature. 

As another byproduct to these developments came a realization that
spin foam models of the Barrett-Crane type are naturally
obtainable from a quantum field theory on a certain group manifold
\cite{DePietri:1999bx}. This generalizes the field theory
introduced by Boulatov \cite{Boulatov:1992vp} in the context of 
3-dimensional quantum gravity. This way of
deriving spin foam models later received a catchy name of ``group
field theory'', see \cite{Freidel:2005qe} for a review. These developments were important at least for
the fact that they allowed to have an efficient bookkeeping tool for the
various amplitudes that are part of the spin foam model (the amplitudes for
the lower-dimensional ``cells'' were not very clearly understood and/or
described in the work of Barrett and Crane \cite{Barrett:1997gw}). 
Most optimistically, the group field theory method  
may have a fundamental status and has a potential to lead to an
independent approach to quantum gravity. 

After this initial period of rapid development the subject of spin
foam models entered a much calmer stage. During this stage evidence was
accumulated that casted serious doubts on the physical
correctness of the Barrett-Crane model(s). Namely, it became clear
that, in spite of correctly capturing at least some aspects of
quantum geometry of a single simplex in 4 dimensions, this theory
is likely not viable as a theory of quantum geometry of more
complicated objects constructed from several such simplices. The
suggested problems with the models were of two main types: 1) the
absence or very limited nature of correlation between amplitudes
of neighboring simplices glued together (the so-called ultra-locality problem); 
2) analytical and numerical evidence (in the case of a single simplex) 
that the simplicial geometries
dominating the partition function of the model in the limit of large 
representations are the degenerate ones \cite{Baez:2002rx}, \cite{Freidel:2002mj},
\cite{Barrett:2002ur}. These problems make it unlikely that the usual world described by a
non-degenerate metric can arise in the semi-classical limit of
these models. Many other criticisms (as well as
some counterarguments to them) of the Barrett-Crane model(s) were
brought forward, and for the past years the subject experienced
some stagnation. 

A new development came with the work by Livine and Speciale \cite{Livine:2007vk},
who used the coherent state methods to rewrite the formulae for the BF theory 
vertex amplitudes. They also suggested that these methods
may prove instrumental in finding an improved version of the
Barrett-Crane model that is free from the above mentioned problems.

Shortly afterwards, there appeared an important paper by
Engle, Pereira and Rovelli \cite{Engle:2007uq}, who 
proposed a new model that was by construction free from
the ultra-locality problem. The authors
of \cite{Engle:2007uq} argue that the key problem
of the Barrett-Crane model(s) is that they impose certain
geometric constraints (the so-called cross-simplicity
constraints) ``strongly'', while some other approaches
to the problem, notably the canonical approach, suggest that these
constraints should be imposed only weakly, if one is not to loose
the degrees of freedom of the gravity theory in question. On a
more technical side, the paper \cite{Engle:2007uq} also hinted at a
new way in which the above mentioned geometric constraints can be
looked at. This new interpretation of the ``simplicity''
constraints was further developed  by Alexandrov in his recent work \cite{Alexandrov:2007pq}.
While the present work was in its final stages, an extended version \cite{Engle2} of 
the paper \cite{Engle:2007uq} has appeared.

The present work resulted from an attempt to understand and
extend the construction of \cite{Engle:2007uq} using methods similar to
\cite{Livine:2007vk}. To this end, we have further developed the new 
geometric perspective on the simplicity constraints, as well
as the new geometrical interpretation suggested by the use
of coherent states in \cite{Livine:2007vk}. As the first immediate
result of our analysis came the realization that the problem
of ultra-locality common to all Barrett-Crane-type models results
from a way these models impose the so-called simplicity constraints.
This way of dealing with the constraints focuses on a single
4-simplex and completely ignores the fact that the amplitude 
associated to this 4-simplex is to become a part of a quantum
amplitude associated to the whole triangulation. The use of coherent
state makes this feature of Barrett-Crane models crystal clear,
as well as suggests a way to remedy this problem. 

As a further result of our analysis came a realization that there is not one,
but two distinct ways to impose the simplicity constraints. The fact that
there should be two different distinct approaches to
``simplicity'' (as suggested by the fact that there are two
distinct symplectic structures on the phase space in question) was
noted long ago, and was one of the early arguments
brought forward against the Barrett-Crane models, which left no
place for a second description. This ambiguity gave grounds
to suspect that the Barrett-Crane models might have given a quantization of
the ``wrong'' sector of the theory. However, later some rather convincing
arguments were put forward to the opposite effect, see in
particular \cite{Baez:1999tk}. The coherent-state-inspired way of
imposing the simplicity constraints leads to a much clearer understanding
of this issue. Thus, we will see that there is not one, but two distinct
models that should be interpreted as giving quantization of the gravitational
and topological sectors of the theory respectively.
One of the models that results from our analysis is that proposed in \cite{Engle:2007uq}.
The other model that we find is new. It is similar to the model of \cite{Engle:2007uq}
in the sense that the simplex amplitudes are strongly correlated and there
is no ultra-locality. However, it also captures some of the main positive features of the 
Barrett-Crane model. The simplex amplitudes for this model are
quite different from the model of Engle et al. \cite{Engle:2007uq}, 
see below for a detailed comparison. Importantly, we show that the model \cite{Engle:2007uq}
corresponds to the topological sector of the theory, while the
new model we propose gives the gravitational sector.
Our arguments are based on a detailed study of the gravitational and topological sectors and on an 
observation that the new model keeps certain key characteristics of the Barrett-Crane model. 
The model we propose thus results by merging  some key features of the Barrett-Crane model with 
the new insight about ultra-locality achieved in \cite{Engle:2007uq}.

The model we obtain (as well as the model of \cite{Engle:2007uq}) 
has two main advantages as compared to the
Barrett-Crane one: First, it clearly possesses a higher
degree of correlations among the neighbouring simplices, thus overcoming
the ultra-locality of the Barrett-Crane model; Second it uses 
$\SU(2)$ structures that bear more resemblance to those 
appearing in traditional loop quantum gravity. Moreover,
the fact that we now have two distinct models that clearly correspond
to two different sectors of the theory -- the gravitational and
the topological ones -- removes the above mentioned shortcoming
of the model of Barrett and Crane that failed to describe the opposite
sector. This availability of the description of both sectors
is what makes the new developments particularly exciting. 

Last but not least, one of the main outcomes of our analysis in the present work 
is that the coherent state methods suggest a way of naturally incorporating 
the so-called Immirzi parameter into the framework of spin foam models,
something that was an open problem for a long time. Indeed, 
the absence of a Barrett-Crane-like model that 
would incorporate the Immirzi parameter was long thought to be a
suspicious feature of the spin foam framework.

We also describe very briefly how the new models
can be obtained within the framework of group field theory. In addition,
our construction allows us to generalize the arguments to the Lorentzian
case and obtain a new model in this setting. The new Lorentzian
model that we obtain has similar advantages as compared to
\cite{Barrett:1999qw}. 

The organization of the paper is as follows. Section \ref{sec:simplicity}
develops a new geometric perspective on the simplicity constraints. 
Section \ref{sec:coherent} introduces the central technique of
coherent states and gives the description of the topological
BF theory model using this technique. We re-derive the Euclidean models 
of \cite{Barrett:1997gw} and \cite{Engle:2007uq} in section \ref{sec:euclid}, 
and also obtain a new model here. Section \ref{sec:imm} describes
how the Immirzi parameter can be introduced into the spin foam model framework. 
We generalize the constructions to the Lorentzian setting in section \ref{sec:lorentz}
and derive a new Lorentzian model. Section \ref{sec:gft} discusses how
the models considered in this paper can be casted into the group
field theory framework. 
In section \ref{geom} we discuss an alternative way to impose the constraints. We finish with conclusions.

After a first version of this paper was placed on the arxiv, a whole set of related works
appeared. In particular, paper \cite{Livine:2007ya} is very close in spirit to the
present work
and paper \cite{Engle:2007wy} studies further the finite Immirzi parameter case. 

\section{A new approach to the geometric simplicity constraints}
\label{sec:simplicity}

Historically, the spin foam models were obtained by applying the
quantization procedure to a geometric simplex in 4 dimensions. One starts 
by associating a Lie-algebra valued variable $X_f$ to each of the 10 faces $f$ 
of the simplex. The Lie-algebra in question is that of the Lorentz group
appropriate to the situation at hand, thus either ${\rm SO}(4)$ in
the Euclidean case or ${\rm SL}(2,\C)$ in the Lorentzian case. A
norm of the Lie-algebra element introduced this way receives the
geometric interpretation of the area of the corresponding face
(note that, in fact, there are two different norms on the
Lie-algebras in question; the duality exchanging these two norms
will play an important role in what follows). The quantization 
procedure then promotes this Lie-algebra element into a quantum 
operator, see more on this below. The spin foam models are then
obtained by analyzing what certain geometrical constraints
appropriate for a flat geometric 4-simplex boil down to
at the level of the operators. Selecting the states that
satisfy the operator version of these constraints one
derives the simplex amplitude.

It turns out to be possible, however, to get the same models
by deriving them from a certain simplicial action that can be
thought of as a precise discrete analog of the gravity one. The
main idea for this comes from the so-called BF
formulation of gravity theories, and from the systematic procedure
of deriving spin foam models from the action principle proposed in
\cite{Freidel:1998pt}. 

\subsection{BF formulation of gravity and path integral quantization}

In Plebanski (or BF, as it was subsequently called) formulation of gravitational theory
\cite{Plebanski}, \cite{Capovilla:1991kx}, \cite{Reisenberger:1998fk},
\cite{De Pietri:1998mb} one starts by the $SO(4)$ (or $SO(3,1)$ in the Lorentzian case) 
BF theory 
\be\label{bf-action}
\int B^{IJ} F(A)_{IJ}
\ee
where $F(A)$ is the curvature of an $SO(4)$  connection and 
$B$ is a  Lie-algebra valued two-form field.
This two form needs to satisfy certain ``metricity'' (or ``simplicity'') constraints
\be \epsilon_{IJKL}B^{IJ}_{\mu\nu}\wedge B^{KL}_{\nu \rho} \propto \epsilon_{\mu\nu \rho \sigma}
  \ee
that guarantee that it is of the form either 
$B^{IJ}=\pm\theta^I\wedge \theta^J$  (called sector I in \cite{De Pietri:1998mb}) or
$B^{IJ}=\pm\epsilon^{IJKL}\theta_K\wedge\theta_L$ (called sector II in \cite{De Pietri:1998mb}).
Here $\theta^I$ are the tetrads, $I,J,K,L$ are ``internal'' indices taking values
 $0,1,2,3$, and
$\epsilon^{IJKL}$ is the completely anti-symmetric tensor. 
Sector II gives the Cartan-Weyl form of Einstein's theory and 
we will refer to it as the gravitational sector. Sector I gives
a theory with no local degrees of freedom. We will call this sector topological.

The path integral quantization of gravity in the BF formulation has been studied
in detail in \cite{Buffenoir:2004vx}. The analysis is complicated by the fact
that the dynamical system is question is quite non-trivial. Indeed, to understand
the correct gauge-fixing terms that need to be added to the action to make sense
of the path integral it is necessary to understand the structure of the symmetry
algebra of the theory, and thus understand the structure of the phase space. Here,
on top of the usual Gauss and the above simplicity constraints as primary constraints
one finds also many secondary constraints (in the topological sector of the theory
there are also tertiary constraints). The resulting constraint algebra has both first-class 
and second-class parts. The first-class constraints are to be gauge-fixed, which converts 
them into second class. The path integral can then be taken over the reduced phase
space. It is important to take into account the fact that the path integral
measure includes a square root of the determinant of the Dirac symplectic structure
on the reduced phase space, and thus may be non-trivial. 
This non-trivial procedure has been carried out in Section IV of \cite{Buffenoir:2004vx},
and the resulting measure on the original space of fields has been found. 

However, what is often not appreciated is that the result of this complicated procedure is 
rather simple. If one, as we are here, is only interested in the partition function,
then one can recast the result of this complicated analysis in terms of the initial 
variables of $BF$ theory subjected to the initial naive primary second class constraints.
The only non-triviality that remains lies in the $B$ field measure factor.
This factor is purely ultralocal, independent of the connection and therefore does not affect 
the analysis or the discretisation of any of the constraints to be imposed. The key point about 
this strategy is that the imposition of the constraints is done at the classical level 
and not at the level of some quantum operator.

This drastic simplification might not occur if one deals not with the partition function,
but with the expectation values of some observables, see e.g. \cite{Alexandrov:2008da} for 
a recent discussion of this point. As argued in \cite{Alexandrov:2008da}, this may strongly affect 
the analysis of the operatorial version of the constraint as performed in \cite{Engle:2007uq, Engle:2007wy}.
However, since in the present work we are interested only in the partition function,
the strategy is to re-write the reduced phase space path integral in terms of the original 
unconstrained variables, together with a set of $\delta$-functions imposing all
the constraints and gauge-fixings, together with a simple ultralocal modification of the path integral measure.

\subsection{Discretized gravity action and the Barrett-Crane model}

The idea behind the spin foam model approach to quantum gravity is to 
appropriately discretise the gravitational action, and perform all
impositions of the constraints, gauge fixings, as well as the path integral quantization
in this discrete setting. It is very well understood how this procedure is realized 
in the context of topological BF theory. It is then the availability of the BF formulation of gravity, 
which makes gravity, at least superficially, look not too distant from BF theory, that makes 
us hope that one day the gravity case will also be understood fully. 

Then our strategy, which is justified by the results of Henneaux et al. \cite{Buffenoir:2004vx} reviwed above,
is to restrict the measure of integration over the classical configurations that enter
into the discretisation of the BF path integral. The results of \cite{Buffenoir:2004vx} tell us that 
one just has to impose the primary second-class constraints - the simplicity constraints.
As usual, there are discretization ambiguities in passing from the continuum to the
discrete picture. However, as we will see, once we have fixed the discrete set of dynamical variables 
and have chosen on which set of variables to impose the constraints,
the implementation of the constraints is rather straightforward and almost unambiguous.
As we shall see, one ambiguity that is left is in the choice of the edge amplitudes and 
amounts to the choice of the ultralocal measure on the $B$ field. Another subtlety lies in the fact 
that we impose the constraint in a linear manner, which involves an introduction of an additional 
normal vector $n_{I}$, see below, and the full constraint analysis of this case has not yet been shown 
to be equivalent to the one of \cite{Buffenoir:2004vx}. In the first archive version of the present 
work we have given strong evidence that the two formulation of the constraints lead to the same 
measure (we have shown it for only half the constraints). For simplicity and coherence 
we have left this analysis out of this printed version, as it is not directly 
relevant to the main line of argument of this work. Another possible worry
about our procedure is whether the canonical construction of the path integral measure 
in the continuum commutes with the discretisation. Thus, a canonical analysis similar to 
\cite{Buffenoir:2004vx} should definitely be attempted at the discrete level. 
In spite of these shortcomings, the procedure that we propose in this paper seems to
be well motivated by the continuum results, and, we hope, provides an important
new step in the direction of the completion  of the  spin foam quantization 
programme.

The above discussion of the ideological foundations of the spin foam method
makes us ready to dive into more technical intricacies. Thus, when discretizing the 
theory on a simplicial complex 
one integrates the two-form $B^{IJ}$ over each  face of the triangulation and 
associates to any face $f$ of the triangulation a Lie-algebra quantity $X_{f}^{IJ}$, see e.g. 
\cite{De Pietri:1998mb}. One also assigns a group element to each tetrahedron and defines 
the following discrete analog of the gravitational action
\be\label{disaction}
S= \sum_{f} \mathrm{Tr}\left(X_{f}G_{f}\right),
\ee
where $G_{f}$ is the holonomy around $f$.
Analogously to the continuous formulation  one has to impose the ``simplicity'' constraints 
on the bivectors $X_{f}$, as well as other second-class together with first-class constraints
on all the discrete variables $X_f, G_f$. However, in this paper we will only
concentrate on the simplicity constraints. Some preliminary ideas on the
other, not considered in this paper second-class constraints, as well
as the path integral measure are contained in the arxiv version of the present
work. However, because of their rather inconclusive character we have decided
not to include that material in this printed version.

The simplicity constraints are easiest to understand by thinking
about a geometric 4-simplex  in $\R^4$ (or Minkowski space $M^{1,3}$). Given
such a simplex one can then associate to each face an ``area'' bivector that we will call
$A_f^{IJ}$ that is obtained by taking the wedge product of two edge vectors. This way one obtains 
four bivectors $A^{IJ}_{tf}$ per each tetrahedron $t$. There are three constraints that this set of 
bivectors must satisfy:
\begin{enumerate}
\item {\bf Simplicity} of each face bivectors :  
\be A^{IJ}_{f} \tilde{A}_{f IJ} = 0,\ee
where $ \tilde{A}^{IJ}=(1/2)\epsilon^{IJKL}A_{KL}$ is the Hodge dual of a bivector.
This constraint implies that $A^{IJ}_{f}$ is ``simple'', i.e., is
of the form $A^{IJ}=u^{[I} v^{J]}$ and defines a geometrical plane. 
\item {\bf Cross simplicity} :  \be A^{IJ}_{f} \tilde{A}_{f' IJ} = 0\ee for two different faces $f,f'$ in $t$.
These constraints guarantee that from the four vectors defined by two simple bivectors $A^{IJ}$ only 3 are 
linearly independent and that therefore the planes defined by them span a three-dimensional subspace in $\R^4$. 
\item {\bf Closure} : \be \sum_{f\subset t} A^{IJ}_{f}=0, \ee 
This constraint fixes the relative coefficient expressing the linear dependence among the bivectors,
which insures that the norms of these bivectors can be interpreted as areas of the faces of a 
geometric simplex.
\end{enumerate}
For more details on these constraints, see for example \cite{Baez:1999tk}. Importantly,
the above constraints are not enough to guarantee that the quadruple of bivectors
is one coming from a geometric tetrahedron, as there are certain discrete ambiguities.
This point is also emphasized in \cite{Baez:1999tk}
and plays an important role in the arguments of this paper.
At the quantum level the bivectors $A^{IJ}_{f}$ are promoted to the Lorentz group 
Lie algebra elements. The above  ``simplicity'' constraints then have to be imposed on these
operators. Barrett and Crane  \cite{Barrett:1997gw} realized that the ``simplicity'' constraints can
be imposed at the quantum level  by selecting  the so-called simple representations
of the Lorentz group, intersection constraints are imposed by
selecting the intertwiner to be the so-called Barrett-Crane
intertwiner, and the closure constraints can be imposed by selecting only 
the gauge-invariant intertwiners. From the point of view of the discrete action
(\ref{disaction}) the variables $X_f^{IJ}$ that appear in the action must
be identified with the Hodge dual $\tilde{A}^{IJ}_f$ of the area bivector $A^{IJ}$. 
Below we shall see how this procedure can be 
implemented.

\subsection{A new geometric criterion for (cross-)simplicity}

In this work we will base our discussion
on a somewhat different interpretation of the simplicity constraints. 
This new interpretation will in particular help us to deal with
the discrete ambiguities that arise. The simplicity constraints,
as reviewed above, are quadratic in the bivector variables, and this
is ultimately the cause of the discrete ambiguities that the set of
solutions possesses. The new formulation of these constraints
makes them linear in the bivectors, thus removing the discrete
ambiguity. Thus, the new geometric way to look at the simplicity and intersection constraints
is contained in the following two lemmas:
\begin{lemma}
A bivector $X^{IJ}$ in $\R^4$ ($M^{1,3}$) is an anti-symmetrized product of two vectors
if and only if there exists a vector $n^I$ such that $X^{IJ} n_J = 0$.
\begin{proof}
If the bivector is simple it defines a two-plane in $\R^4$ ($M^{1,3}$). Taking $n^I$ to
be any of the vectors orthogonal to this plane proves the assertion in one direction.
To prove it in the opposite direction we assume that $X^{IJ}$ is not simple. Then
there always exist 4 vectors $u_{1,2}, v_{1,2}$ so that
\be
X^{IJ} = u_1^{[I} v_1^{J]} +  u_2^{[I} v_2^{J]},
\ee
and  $u_{1,2}, v_{1,2}$ span $\R^4$ ($M^{1,3}$). Therefore,
these vectors can be used as a basis, and the vector $n^I$ can
be represented as a linear combination of these basis vectors.
It is then easy to see that the condition $X^{IJ} n_J=0$ says
that the 4 vectors  $u_{1,2}, v_{1,2}$ are linearly dependent.
This contradicts our assumption that $X^{IJ}$ is not simple.
\end{proof}
\end{lemma}

\begin{lemma} Two simple bivectors $X_1^{IJ}$ and $X_2^{IJ}$
span a 3-dimensional subspace of $\R^4$ ($M^{1,3}$) if and only if
there exists a vector $n^I$ such that $X_1^{IJ} n_J = 0$ and
$X_2^{IJ} n_J = 0$.
\begin{proof}
If two bivectors are simple and the 2-planes defined by them span
a 3-dimensional subspace, then $n^I$ can be chosen to be a vector
orthogonal to this subspace. This proves the assertion in one direction.
The proof in the opposite direction is similar to the proof
in the above lemma in that one shows that the condition of
existence of $n^I$ implies that the 4 vectors defined by the
2 two-planes must be linearly dependent, and thus span a
3-dimensional subspace.
\end{proof}
\end{lemma}

Importantly, the above simplicity and intersection constraints can
also be formulated in exactly the same way in terms of the Hodge
duals $ \tilde{X}^{IJ}=(1/2)\epsilon^{IJKL}X_{KL}$ of the bivectors.
Thus, we have the following two ``dual'' lemmas:
\begin{lemma}
A bivector $X^{IJ}$ in $\R^4$ ($M^{1,3}$) is is an
anti-symmetrized product of two vectors if and only if there
exists a vector $n^I$ such that $ \tilde{X}^{IJ} n_J = 0$.
\begin{proof}
A bivector $X^{IJ}$ is simple if and only if its dual bivector
$\tilde{X}^{IJ}$ is simple, so the first lemma above also proves
this lemma.
\end{proof}
\end{lemma}

\begin{lemma} Two simple bivectors $X_1^{IJ}$ and $X_2^{IJ}$
span q 3-dimensional subspace of $\R^4$ ($M^{1,3}$) if and only if
there exists a vector $n^I$ such that $\tilde{X}_1^{IJ} n_J = 0$ and
$\tilde{X}_2^{IJ} n_J = 0$.
\begin{proof}
Two bivectors $X_{1,2}^{IJ}$ span a 3-dimensional subspace if and
only if their duals $\tilde{X}^{IJ}_{1,2}$ do, so the second lemma
above proves this lemma as well.
\end{proof}
\end{lemma}

An important point is that, in spite of having these
two seemingly equivalent ways to impose the simplicity
and intersection constraints, they are no longer equivalent
when one starts to deal with more than two bivectors.
Thus, considering, as in the previous subsection, the
area bivectors $A_f^{IJ}$ obtained by taking the wedge
product of the edge vectors corresponding to a tetrahedron
one can see that $A_f^{IJ}$ satisfy the simplicity and
intersection constraints in their first version, namely
that there exists a vector $n^I$ orthogonal to all $A_f^{IJ}$,
but not in the second version.
 Indeed, let us consider a 3-dimensional
tetrahedron built on unit vectors along the axes $1,2,3$.
The corresponding area bivectors are: $A^{IJ}_{12}=e^I_1\wedge e^J_2$,
$A^{IJ}_{23}=e^I_2\wedge e^J_3$, $A^{IJ}_{31}=e^I_3\wedge e^J_1$, and
the fourth bivector being $-A^{IJ}_{12}-A^{IJ}_{23}-A^{IJ}_{31}$. 
Here $e^I_{0,1,2,3}$ are  unit vectors in the direction of the axes.
All 4 bivectors are orthogonal to the vector $e^I_0$. However, the
dual bivectors $\tilde{A}^{IJ}_{12}=e^I_0\wedge e^J_3, 
\tilde{A}^{IJ}_{23}=e^I_0\wedge e^J_1, \tilde{A}^{IJ}_{31}=e^I_0\wedge e^J_2$
{\it do not} satisfy the simplicity-intersection constraints in their
first version, as there is no vector such that 
$\tilde{A}^{IJ}_{12}, \tilde{A}^{IJ}_{23}, \tilde{A}^{IJ}_{31}$
would all be orthogonal to. Instead, these bivectors satisfy
the simplicity constraints in their second version, they are all parallel to $e_{0}$. 

It is therefore clear that the simplicity and cross-simplicity
constraints can be imposed on a set of 4 Lie-algebra valued variables
$X^{IJ}_f$ in two ``dual'' ways. Indeed, the constraints become the requirement
that there exist a vector $n^I$ such that the 4 variables $X^{IJ}_f$ are either all orthogonal
$X^{IJ}_f n_{t\, J} = 0$  to the vector $n^I_t$ or are all parallel  
$\tilde{X}^{IJ}_f n_{t\, J}=0$  to it. This is the form of the simplicity constraints that we will use.

In the first case $X^{IJ}_f n_{t\, J} = 0$ the
variables $X^{IJ}_f$ can be interpreted as the area bivectors 
of a geometric tetrahedron: $X^{IJ}_f=A^{IJ}_f$, while
in the second case $\tilde{X}^{IJ}_f n_{t\, J}=0$ the duals $\tilde{X}^{IJ}_f$ admit
the same interpretation: $\tilde{X}^{IJ}_f=A^{IJ}_f$.

\subsection{Self- and anti-self-dual decomposition}

Consider now one of the variables $X^{IJ}$. Consider the following
two quantities: 
\be X^{\pm J}(X):=\left( n_{I} \tilde{X}^{IJ}\pm\sqrt{\sigma} n_{I}X^{IJ} \right). 
\ee 
Here $n^I$ is at this stage arbitrary unit vector (future timelike in the Lorentzian case) and 
$\sigma$  is $1$ in the Euclidean signature and $-1$ in the Lorentzian case.
It is easy to see that $X^{\pm I}n_I=0$, so both vectors $X^{\pm}$ actually
take values in the 3-dimensional subspace orthogonal to $n^I$. It
is also easy to check that 
\be X^{\pm J}(\tilde{X}) = \pm \sqrt{\sigma}  X^{\pm\,I}(X), \ee 
where $\tilde{}$ is the Hodge duality operator. 
Thus, $X^{\pm}$ are eigenvectors of this operator.
They are called the self- and anti-self-dual parts of $X^{IJ}$.
Their knowledge is equivalent to the knowledge of $X^{IJ}$ (there
is an explicit reconstruction formula that we will not need here).

As we have seen above, when $X^{IJ}$ is simple there exists a vector $n^I$ such that
$X^{IJ} n_J=0$ or, equivalently, a different vector such that $\epsilon^{IJKL}n_J X_{KL}=0$. 
The two distinct criteria of simplicity (and intersection) can then be formulated in 
terms of self- and anti-self-dual parts of bivectors as:
\beq\label{1}
\mathrm{I:}\quad& X^{IJ} n_J = 0 &\Longleftrightarrow 
X^{+ J} = X^{- J} = n_{I}\tilde{X}^{IJ},\\ \label{2}
\mathrm{II:}\quad&   \epsilon^{IJKL} n_J X_{KL}=0  &\Longleftrightarrow  
X^{+ J} = -X^{- J} = \sqrt{\sigma} \,  n_{I} X^{IJ}. 
\eeq

\subsection{Identification of bivectors with Lie algebra elements}

From now on we restrict our discussion to the Euclidean case. As we have already 
said, the idea of quantization is to first associate to each bivector a Lie algebra element
\be
\hat{\bf{X}}= X^{IJ}\hat{J}_{IJ},
\ee
where $\hat{J}_{IJ}$ are the generators of $SO(4)$
\be
[\hat{J}_{IJ}, \hat{J}_{KL}] = i \delta_{IK} \hat{J}_{JL} - i \delta_{JK} \hat{J}_{IL} - 
i \delta_{IL} \hat{J}_{JK} + i \delta_{JL} \hat{J}_{IK}.
\ee
Let us now pick a fixed four vector $n_{0}=(1,0,0,0)$ and define the corresponding 
self-dual operators:
$ X^{\pm i} =( X^{i} \pm X^{0i})$, where $X^{i}\equiv 1/2 \epsilon^{i}{}_{jk} X^{jk}$ and
$ \hat{J}^{\pm}_{i} =( \hat{J}_{i} \pm \hat{J}_{0i})$.
The Lie algebra element then decomposes in terms of its self- and anti-self-dual parts
\be
 \hat{\bf{X}} = \hat{X}_{+} +\hat{X}_{-}, \quad \hat{X}^{\pm} =X^{\pm i}\hat{J}^{\pm}_{i}.
\ee
The signs entering the definition of the self-dual operators are such that 
they form an $su(2) \oplus su(2)$ Lie algebra
\be
[\hat{J}^{\pm}_{i},\hat{J}^{\pm}_{j}] =i 2 \epsilon_{ij}{}^{k}\hat{J}^{\pm}_{k},\quad [\hat{J}^{+}_{i},\hat{J}^{-}_{j}] = 0.
\ee
At the group level the described splitting of the Lie algebra into the self- and anti-self-dual parts
leads to the isomorphism ${\rm SO}(4) = {\rm SU}(2)\times{\rm SU}(2)/\Z_2$, so that
each ${\rm SO}(4)$ group element $\bf{g}$ can be represented as ${\bf{g}}=(g^+,g^-),
 g^\pm\in{\rm SU}(2)$. This isomorphism can be described more explicitly as follows.
A point in $\R^4\sim \C^2$ can be parametrized by a pseudo-unitary $2\times 2$ matrix
\beq
x=\left( \begin{array}{cc} \alpha & \beta \\ -\bar{\beta} & \bar{\alpha} \end{array} \right).
\eeq
Transformations
\be\label{embedding}
\SU(2)\times\SU(2) \ni (g^+, g^-): x\to g^+ x (g^-)^{-1}
\ee
preserve the determinant ${\rm det}(x)=|\alpha|^2+|\beta|^2$ and are thus orthogonal
transformations of $\R^4$. Of particular interest for us is the ``diagonal'' subgroup of
$\SU(2)\times\SU(2)$ given by elements of the form $(g,g)$. Let us also mention that
under the adjoint action of $SO(4)$ the self- and anti-self-dual elements transform 
as $SU(2)$ Lie algebra elements:
\be
{\bf g} {\bf X} {\bf g}^{-1} = g^{+}\hat{X}^{+}(g^{+})^{-1} + g^{-}\hat{X}^{-}(g^{-})^{-1}.
\ee

\subsection{Gravitational and topological sectors}

Now, the key point for us is that, as a comparison between the continuous (\ref{bf-action}) and 
discrete (\ref{disaction}) actions suggests, if the theory is to correspond to the gravitational
sector the quantities $X^{IJ}_f$ in (\ref{disaction}) should be identified with the {\it duals} of 
the area bivectors $A_f^{IJ}$:
\be\label{X-A}
X_{f}^{IJ} =\frac{1}{2} \epsilon^{IJ}{}_{KL} A_{f}^{KL}.
\ee
Correspondingly, the simplicity and intersection constraints should be imposed in their
version (\ref{2}). The other case, namely with $X_f^{IJ}$ satisfying the
constraints in their version (\ref{1}) corresponds to the topological sector. In
the subsequent sections we will see that there exist two corresponding models,
one for the gravitational and one for the topological sector.

Let us rewrite the criteria (\ref{1}), (\ref{2}) in a more convenient form. We will
make all statements about the gravitational sector case. The other case is
obtained by a simple duality transformation. Thus, as we have just seen, given four face bivectors 
$X_{f}$ associated with a tetrahedron $t$ and satisfying the 
simplicity and cross simplicity (intersection)
conditions there exist a unit vector $n_{t}$ such that  $n_{t\,I} \tilde{X}^{IJ}_{f} =0$. There
always exists an ${\rm SO}(4)$ transformation that brings the vector $n_t^I$ into the
fixed reference vector $n^I_{0}$. Let us denote the inverse of this transform by $(g^+,g^-)$. Thus,
$n_{0}=1 \to n= g^{+}(g^{-})^{-1}$. We have seen that in the case when $n^I$ is chosen to point
along the ``time'' axis $n^I=(1,0,0,0)$ the self-and anti-self-dual part of a bivector
satisfying the simplicity conditions in their form (\ref{2}) satisfy $X^{+}_{f}=-X^{-}_{f}$ so that 
${\bf X}_{f}=(X_{f},- X_{f})$. The ${\rm SO}(4)$ rotation maps this bivector into 
\be (X_{f},- X_{f})\to (g^{+}X_{f}(g^{+})^{-1}, - g^{-}X_{f}(g^{-})^{-1}). \quad \ee
This makes it clear that in the case of a general normal vector $n_t$ the simplicity-intersection
constraints on $X_{f}$ corresponding to the gravitational sector can be formulated as:
\be\label{grav}
  {\bf X}_{f} = ( \hat{X}_{f}, -n_{t}^{-1}\hat{X}_{f}n_{t}),\ee
 where $n_{t}\in{\rm SO}(4)$ should be interpreted as carrying 
information about the normal vector to the 
tetrahedron containing $X_f$. Similarly, the constraints appropriate 
to the topological sector take the form:
  \be\label{top}
  {\bf X}_{f} = (  \hat{X}_{f}, n_{t}^{-1}\hat{X}_{f}n_{t}).\ee
  These are the final criteria we will use in the following.

In the next section
we would like to impose the conditions (\ref{grav}) and (\ref{top}) at the quantum level.
We will achieve this via the technique of the coherent states introduced in the context of spin foams by 
Livine and Speziale \cite{Livine:2007vk} and obtain two
distinct models that we will interpret as corresponding to the gravitational
and topological sectors. The technique of coherent states that we shall use will also allow us to see
that the model of Barrett and Crane does impose the correct gravitational  constraints 
(\ref{grav}) but in a form that is too strong: the constraints are imposed for
each  4-simplex independently of the others simplices of the triangulation. 
This will be interpreted
as a likely source of all the problems the Barrett-Crane model is known to suffer from.  
The model of Engle, Pereira and Rovelli \cite{Engle:2007uq} on the other hand
will be seen to impose the simplicity constraints in a satisfactory manner. However, 
as we will see, this model turns out to give a quantization of the topological sector
in that it imposes the constraints in their version (\ref{top}), and is thus unsuitable as a model 
for quantum gravity. A new model that imposes the constraints in their version (\ref{grav})
will be obtained below.

\section{Coherent states and the spin foam model of BF theory}
\label{sec:coherent}

The technique of coherent states is crucial for all the constructions that follow. We will
therefore start by a review of this technique. Following \cite{Livine:2007vk}, we will
illustrate this technique by applying it to the BF theory partition function.

\subsection{Coherent states}

Let us first describe the coherent states for the
group ${\rm SU}(2)$. It will then be trivial to generalize
it to ${\rm SO}(4)$. 

We will start from the following obvious decomposition of the identity
operator in the representation space $V^j$ of dimension $\rd_j\equiv 2j+1$:
\be\label{ident-m}
1_j = \sum_m |j,m \ket \bra j,m|,
\ee
where $|j,m\ket, m\in[-j,j]$ is the usual orthonormal basis in $V^j$.
Let us rewrite the same operator by inserting in it the Kronecker delta
\be
\delta_{mm'} = \rd_j \int_{{\rm SU}(2)} dg \, t^j_{mj}(g) \overline{t^j_{m'j}(g)}
\ee
written in terms of an integral over the group. Here $t^j_{mj}(g)$ is the
matrix element of the group element $g$ in the representation $j$ computed between
the states $\bra j,m|$ and  $|j,j\ket$, the latter being the highest weight state. More
generally:
\be
t^j_{mm'}(g) \equiv \bra j,m| g|j,m'\ket.
\ee
This gives:
\be\label{ident-coherent}
1_j = \rd_j \sum_{mm'} |j,m \ket \bra j,m'|
\int_{{\rm SU}(2)} dg \, t^j_{mj}(g) \overline{t^j_{m'j}(g)} =
\rd_j \int_{{\rm SU}(2)} dg \, |j,g \ket \bra j,g|,
\ee
where we have introduced a notation:
\be\label{def}
|j,g \ket \equiv g |j,j\ket = \sum_m  |j,m \ket t^j_{mj}(g) .
\ee
The states $|j,g \ket$ are known as the coherent states, and
the last expression in (\ref{ident-coherent}) is the decomposition
of the identity in terms of these coherent states.

The decomposition of the identity (\ref{ident-coherent}) can be further simplified by 
noticing that the matrix elements $t^j_{mj}(g)$ and $t^j_{mj}(gh)$ differ only by a phase for 
any group element $h$ from the ${\rm U}(1)$ subgroup of ${\rm SU}(2)$. Therefore, the integral
in (\ref{ident-coherent}) can be taken over the coset $G/H, G={\rm SU}(2), H={\rm U}(1)$. Thus, we have:
\be
1_j = \rd_j \int_{G/H} dn \, |j,n \ket \bra j,n|.
\ee
We will suppress the domain of integration $G/H$ in what follows.

The states $|j,n\ket$ form (an over-complete) basis in $V^j$. 
One of the main motivations for using these states comes from the fact that they have  
a clear geometrical interpretation: Let $\hat{J}^{i}$ be the generators of the $SU(2)$ 
Lie algebra $[\hat{J}^{i},\hat{J}^{j}]= i \epsilon^{ijk} \hat{J}_{k}$. We can easily compute 
the expectation value of these generators in a given coherent state:
\be \label{geo}
X_{(j,n)}^i \sigma_i \equiv \bra j,n|\hat{J}^{i}|j,n \ket \sigma_{i} =  j \, n\sigma_{3} n^{-1}
\equiv j \,n^{i} \sigma_{i},
\ee
where the quantities $\sigma_{i}$ are the Pauli matrices and 
we have denoted by the same letter the $SU(2)$ matrix $n$ that parametrized the coherent state 
and the three-dimensional unit vector $\vec{n}$ that this matrix corresponds to.
Thus we see that  $|j,n\ket$ describes a vector in $\mathbb{R}^{3}$ of length $j$ and of direction 
determined by the action of $n$ on a unit reference vector. It is convenient to denote this vector by the 
same letter $\vec{n}$. This geometric interpretation will be extremely useful for us in what follows.

An important remark is as follows. Note that we could in principle define a notion of ``coherent'' state 
starting from any state $|j,m\ket$ instead of the highest weight vector $|j,j\ket$. Indeed, we
would have the same decomposition of unity as in (\ref{ident-coherent}). The geometrical interpretation would
be, however, drastically different. Namely, we could no longer interpret the states $g |j,m\ket$ as representing
a vector lying on the sphere of radius $j$. Instead, these states correspond to a circle on this sphere of 
radius $\sqrt{j^{2}-m^{2}}$. Moreover, we can compute the dispersion of the quadratic operator $\hat{J}^{2}$ 
with the result being \be \Delta J^{2} = j + j^{2}-m^{2}.\ee
Thus, we see that only the highest and lowest states $m=\pm j$ lead to the coherent states that minimize the 
uncertainty relation.

It will be of crucial importance for us that there is not one but two coherent states $g |j,j\ket\equiv |j,g\ket$ and 
$g|j,-j\ket\equiv \overline{|j,{g}\ket}$. In order to relate them let us remark that the group element 
\be
\epsilon \equiv  \left( \begin{array}{cc} 0 & 1 \\ -1& 0 \end{array} \right)\ee
is such that 
\be 
g\epsilon = \epsilon \bar{g},
\ee
where $\bar{g}$ is the complex conjugate matrix.
The matrix element of $\epsilon$ in any representation is given by 
\be t_{nm}^{j}(\epsilon)= (-1)^{j-n} \delta_{n,-m}.\ee Thus, we have $\bra j,-j| \epsilon|j,j\ket= (-1)^{2j}$ and therefore
the lowest weight vector can be obtained by the action of $\epsilon$ on $|j,j\ket$. We have: 
\be
|j,-j\ket = (-1)^{2j}\epsilon |j,j\ket.
\ee 
More generally we can define the conjugate states
\be
\overline{|j,m\ket} \equiv \epsilon |j,m\ket = (-1)^{j+m}|j,-m\ket.
\ee
These are called conjugate since their matrix elements are the complex 
conjugates of the usual matrix elements
\be 
\overline{\bra j,m|}g\overline{|j,n\ket} = 
\bra j,m|\epsilon^{-1}g\epsilon{|j,n\ket} = \overline{t_{mn}(g)} = (-1)^{2j+m+n}t_{ -m -n}(g)
\ee
We thus obtain two families of coherent states $|j,g\ket$ and 
\be
\overline{|j,{g}\ket} \equiv (-1)^{2j} g |j,-j\ket =  \sum_m    \overline{|j,m \ket} \overline{t^j_{mj}(g)}.
\ee

We are eventually interested in coherent states for ${\rm SO}(4)$. What this discussion shows is 
that one has four possible coherent states which are given by acting with an ${\rm SO}(4)$ group element 
on either of the following states: 
\be 
|j,j\ket\otimes |j,j\ket, \quad|j,j\ket \otimes |j,-j\ket, 
\quad|j,-j\ket \otimes |j,j\ket, \quad|j,-j\ket \otimes |j,-j\ket.
\ee
Note that these four states can be obtained from one another by the action of 
an $SO(4)$ group element. However, if one considers only the action of the diagonal $\SU(2)$ 
subgroup of elements of the form $(g,g)$ then there are two inequivalent 
states that cannot be related by such a transformation. As we will see below the state 
$|j,g\ket\otimes |j,g\ket$ will play the key role in the constructions of the topological sector 
whereas the state $|j,g\ket\otimes \overline{|j,{g}\ket}$ will be the one relevant to the gravitational sector.

\subsection{BF theory}
In the approach to spin foam models that derives them from an action principle, the
starting point is the fact that gravity can be written as a constrained BF theory. 
Thus, the strategy devised in \cite{Freidel:1998pt} is to first construct the partition 
function  of the BF theory on a given triangulation and then impose the simplicity 
constraints at the level of the partition function. Note that this is rather different from 
the strategy followed by Barret and Crane who imposed the simplicity constraints at the level
of each 4-simplex, with the starting point being the idea of quantization of a 4-simplex.
In simplicial version of BF theory, the partition function is obtained by imposing the 
flatness condition on a discrete connection --- using a product of
$\delta$-functions imposing the constraints that the holonomies around
all dual faces are trivial. These constraint are obviously too strong
for a gravitational theory and are to be relaxed. 

To write these constraints more explicitly, we introduce the following
convenient notations. Let us consider two neighboring 4-simplices $\sigma,\sigma'$  of a
simplicial decomposition of a 4-manifold. These two simplices intersect along a tetrahedron 
$t=(\sigma,\sigma')$. One chooses ``centers'' of these
4-simplices (these centers become vertices of the dual complex),
and connects them with an interval - the edge $\sigma\sigma'$ of the dual
complex. This dual edges  are in  one-to-one correspondence with a
tetrahedron $t$ of the original simplicial decomposition.  Consider now
the holonomy $g_t\equiv g_{\sigma\sigma'}$ of the gravitational spin connection along the dual
edge $\sigma\sigma'$. Assuming that the bundle is trivial (or choosing a local trivialization), 
this holonomy is  an element of the group ${\rm SO}(4)$ in the Euclidean case and
${\rm SL}(2,\C)$ in the Lorentzian setting. Taking a product of
these group elements for all the dual edges that form the boundary
of the dual face (one has to introduce an orientation of the dual
face in an arbitrary fashion) one obtains the holonomy around the dual face.
The flatness condition is imposed by requiring all these dual face
holonomies to be trivial. The partition function of the theory is then
given by:
\be\label{z-bf-1}
Z= \int \prod_{t}dg_{t}\prod_{f}\delta\left( \overrightarrow{\prod_{ t\supset f}} g_{t} \right),\ee
The integral here is over group elements associated with each tetrahedron. The $\delta$-functions impose
the constraint that the holonomy around each dual face is trivial.

One then uses the Plancherel decomposition to rewrite the $\delta$-functions 
into a sum over irreducible representations of the group in question. 
In the case of ${\rm SO}(4)$ we use the notation ${\bf g}=(g_{+},g_{-})$ for an 
${\rm SO}(4)$ group element, ${\bf j} =(j_{+},j_{-})$ for an ${\rm SO}(4)$ irreducible representation, 
$\mathrm{d}_{\bf j} = \mathrm{d}_{j_{+}} \mathrm{d}_{j_{-}}$, $\mathrm{d}_{j}=2j+1$ for the corresponding
dimension and  $\mathrm{tr}_{{\bf j}_{f}}({\bf g})= \mathrm{tr}_{ j^{+}_{f}}(g^{+}) \mathrm{tr}_{j^{-}_{f}}(g^{-})$ 
for the characters. It proves instrumental to split each group element $g_t$ in (\ref{z-bf-1}) into
a product of two group elements. Given that the group elements $g_t$ have the interpretation of the
holonomy along the dual edges, they can be represented as a product of the holonomy $g_{\sigma t}$ from
the center of 4-simplex $\sigma$ to the center of tetrahedron $t$ with the holonomy from $t$ to $\sigma'$.
Thus, $g_t=g_{\sigma t}g_{t\sigma'}$. The introduced group elements can be integrated over
separately without changing the partition function (we assume that the group integration measure
is normalized). We will also use the convention that $g_{t\sigma}= g_{\sigma t} ^{-1}$. 
The Plancherel decomposition, together with the introduced trick of doubling the group
elements to be integrated over, reduce the BF partition function to the following expression: 
\be\label{z-bf-2}
Z= \sum_{{\bf j}_{f}} \prod_{f}\mathrm{d}_{{\bf j}_{f}}
\int \prod_{(t,\sigma)}\mathrm{d}{\bf g}_{t\sigma}\prod_{f}\mathrm{tr}_{{\bf j}_{f}}
\left( \overrightarrow{\prod_{ \bra\sigma\sigma'\ket \supset f}} {\bf g}_{\sigma t }{\bf g}_{t\sigma' } \right).
\ee

The key step proposed in \cite{Livine:2007vk} is to insert into this expression a multiple
decomposition of the identity written in terms of the coherent states. 
Thus, for each tetrahedron $t$  
and each dual face we insert into (\ref{z-bf-2}) the following decomposition of unity:
\be
1_{\bf{j}} =\mathrm{d}_{\bf{j}} \int \mathrm{d}{\bf n} 
| {\bf{j}}, {\bf n}\ket \bra {\bf{j}}, {\bf n}|,\quad  | {\bf{j}}, {\bf n}\ket\equiv  
| {{j_{+}}}, { n_{-}}\ket\otimes  | {{j_{-}}}, { n_{-}}\ket.
\ee
This gives 
\be\label{ZBF}
Z= \sum_{{\bf j}_{f}} \prod_{f}\mathrm{d}_{{\bf j}_{f}}\int \prod_{(t,\sigma)}\mathrm{d}{\bf g}_{\sigma t}
\prod_{(t,f)}\mathrm{d}_{\bf{j}_{f}} \mathrm{d}{\bf n}_{t f}
 \prod_{(\sigma,f) }\bra {\bf j}_{f}, {\bf n}_{tf}|  ({\bf g}_{ \sigma t})^{-1}{\bf g}_{\sigma  t'}| {\bf j}_{f}, {\bf n}_{t'f}\ket
\ee
For a triangulation of a 4-dimensional manifold there are exactly 4 dual faces that 
share each particular dual edge, and thus each group element $g_{t\sigma}$ 
enters into 4 characters. Integrating over these holonomies one produces a product 
of 15j symbols -- one per 4-simplex -- and gets what is known as the spin foam model of BF theory 
\cite{Ooguri:1992eb}. This derivation is usually done without the insertion of a complete set 
of states for each tetrahedron, which only complicates the matters in this simple case. However, 
below we will see that the expression (\ref{ZBF}) 
that we wrote is a very useful starting point for the non-topological models.

Another useful interpretation that we can give the expression (\ref{ZBF}) is to think about
the triangulated manifold amplitude that it gives as the product of ``vertex'' amplitudes
corresponding to 4-simplices $\sigma$ (or vertices of the dual triangulation) and ``edge'' amplitudes
corresponding to tetrahedra $t$ (or edges of the dual triangulation). The vertex amplitude
is a function of the representations $\{ j_f\}$ labelling the faces of $\sigma$ as well
as $5\times 4$ group elements ${\bf n}_{tf}$ labelling the faces $f$ of tetrahedra $t$.
Note that the same face $f$ is shared by exactly two tetrahedra, which we denote by $t(f)$ and $t'(f)$.
The vertex amplitude is given by:
\be\label{vertex-BF}
A^\sigma_{BF}[\{ {\bf{j}}_{f} \}, \{ {\bf{n}}_{tf} \}] = \prod_{f}\mathrm{d}_{{\bf j}_{f}} \int \prod_t \mathrm{d}{\bf g}_{t}  \prod_f 
\bra {\bf j}_f, {\bf n}_{t(f)f}| ({\bf{g}}_{t(f)})^{-1} {\bf g}_{t'(f)} |{\bf j}_f, {\bf n}_{t'(f)f} \ket.
\ee
The edge amplitude that corresponds to (\ref{ZBF}) is simply a $\delta$-function that sets
the variable ${\bf n}_{t(f)f}^{\sigma}$ of one 4-simplex $\sigma$ to be equal to a similar variable
${\bf n}_{t'(f) f}^{\sigma'}$ corresponding to the same face $f$ but in another 4-simplex. 
Thus, the edge amplitude is:
\be\label{edge-BF}
A^{tf}_{BF}[\{ {\bf{j}}_{f} \}, \{ {\bf{n}}_{tf}^{\sigma} \}] = \delta({\bf n}_{t(f)f}^{\sigma}, {\bf n}_{t'(f)f}^{\sigma'})
\,\,\,\mathrm{or\, equivalently}\,\,\,
= \mathrm{d}_{{\bf j}_{f}} \bra{\bf j}_{f}, {\bf n}_{t(f)f}^{\sigma}|{\bf j}_{f},{\bf n}_{t'(f)f}^{\sigma'}\ket.
\ee
The full partition function is then obtained by taking a product of vertex and edge amplitudes, 
summing over ${\bf j}_{f}$ and integrating over ${\bf{n}}_{tf}^{\sigma}$.

\section{Imposing the simplicity constraints}
\label{sec:euclid}

In the previous section we have seen that it is possible to associate to each state 
$| {\bf j}_{f}, {\bf n}_{tf}\ket$ a bivector 
\be
{ X}_{( {\bf j}_{f}, {\bf n}_{tf})}^{IJ}=  
\bra {\bf j}_{f}, {\bf n}_{tf}| \hat{ J}^{IJ} | {\bf j}_{f}, {\bf n}_{tf}\ket,\quad \quad 
\hat{\bf X}_{( {\bf j}_{f}, {\bf n}_{tf})}= 
\left( \hat{X}^{+}_{(j_{f}^{+},n_{tf}^{+})}, \hat{X}^{-}_{(j_{f}^{-},n_{tf}^{-})}\right),
\ee
where $\hat{J}^{IJ}$ denotes a basis of  ${\rm SO}(4)$ Lie algebra elements.
According to the results of section \ref{sec:simplicity}, demanding that these bivectors satisfy 
the geometrical simplicity and cross simplicity conditions i), ii) is equivalent to the requirement that
there exists an $\SU(2)$ group element $u_{t}$ (interpreted as characterizing the normal to the 
3-dimensional plane spanned by the tetrahedron $t$) such that 
\be  \hat{\bf X}_{( {\bf j}_{f}, {\bf n}_{tf})} = (  \hat{X}_{f}, - u_{t}\hat{X}_{f}u_{t}^{-1}).\ee 
This condition is the one relevant for the gravitational sector. The topological
sector corresponds to a similar condition, but without the minus sign
in the anti-self-dual part.

Using the fact that $X_{(j_{f},n\epsilon)} = -X_{(j_{f},n)}$ and the fact that 
$X_{(j_{f},nh_{\phi})} = X_{(j_{f},n)}$ if  $h_{\phi}\equiv \exp(i\phi \sigma_{3})$ 
is a ${\rm U}(1)$ element,
we can rewrite the
simplicity conditions for the gravitational sector in terms of the spins $j_{f}$ and the $SU(2)$ elements $n_{f}$ as:
\be\label{grav1}
j^{{+}}_{f} = j^{{-}}_{f},\quad {\mathrm{and}}\quad (n_{tf}^{+}, n_{tf}^{-})=
 (n_{tf} h_{\phi_{tf}},u_{t}  n_{tf} h_{\phi_{tf}}^{-1} \epsilon),
\ee
where $\phi_{tf}\in [0, 2\pi]$. The reason why we have allowed for the presence
of an additional ${\rm U}(1)$ comes because the group elements $n^\mp_{tf}$ 
parametrizing the normals are defined modulo the right ${\rm U}(1)$. 
Note that the bivectors are initially invariant under 
$U(1)\times U(1)$. However, the diagonal $U(1)$ invariance is 
reabsorbed into the definition of $n_{f}$ and the anti-diagonal one is taken care by $\phi_{tf}$. 
We will see below that these angles can play an interesting role in the construction.

Similarly, the simplicity constraints for the topological sector are equivalent to
the following two conditions:
 \be\label{top1}
j^{{+}}_{f} = j^{{-}}_{f}=j_{f},\quad {\mathrm{and}}\quad (n_{tf}^{+}, n_{tf}^{-})=
 (n_{tf} h_{\phi_{tf}},u_{t}  n_{tf} h_{\phi_{tf}}^{-1} ),
\ee
with the only difference being in the absence of the $\epsilon$ group element in the 
anti-self-dual copy. 

\subsection{The Barret-Crane model}

The logic now is to impose the constraints (\ref{grav1}) and (\ref{top1}) 
in the expression (\ref{ZBF}) for the partition function of the BF theory,
with the motivation being that a similar imposition at the level of
the continuous theory is known to produce gravity. It is clear that when
doing this, one will be imposing the constraints at each tetrahedron. Importantly,
as the edge amplitude (\ref{edge-BF}) requires, the constraints are
not imposed at each 4-simplex individually since a tetrahedron is shared by two 
4-simplices. However, let us see what
happens if we ignore this requirement and do impose the constraints
at the level of each 4-simplex. Thus, what we would like to do is 
to impose the gravitational constraint (\ref{grav1}) on the group elements 
$n^{\pm}_{tf}$ entering the definition of (\ref{vertex-BF}) and then integrate 
over $n_{tf}, \phi_{tf}, u_{t}$. The integral over $u_{t}$ can be reabsorbed in a 
redefinition of $g^{-}_{\sigma t}$ and the integral over $\phi_{tf}$ also drops out 
because of the simplicity constraints $j^{+}=j^{-}$. It is interesting to note that even if one does not impose 
the constraint $j^{+}=j^{-}$ it will be automatically implemented by the integral over the 
${\rm U}(1)$ group element $\phi_{tf}$. This gives another reason to introduce this 
group element -- to give a simple tool to
impose the constraints $j^+=j^-$. After all this operations are done, one
is left with the vertex amplitude that only depends on $10$ spins $j_{f}$ and
is given by:
\beq\label{vertex-BC}
A^\sigma_{BC}[\{ {{j}}_{f} \}] &=& \int \prod_t \mathrm{d}{ g}_{t}^{+}\mathrm{d}{ g}_{t}^{-} \prod_{tf}\mathrm{d}n_{tf} \times \\
& &\prod_f  \bra { j}_f, { n}_{t(f)f}| ({{g}}^+_{t(f)})^{-1} { g}^+_{t'(f)} |{ j}_f, { n}_{t'(f)f} \ket
\overline{\bra { j}_f, { n}_{t(f)f}| ({{g}}^-_{t(f)})^{-1} { g}^-_{t'(f)} |{ j}_f, { n}_{t'(f)f} \ket}\nonumber,
\eeq
where we have used the fact that $|j,g\epsilon\ket = \overline{|j,g\ket}$.
The integrals over $n_{tf}$ can be easily taken as follows:
\beq
\int \mathrm{d}n\, |j,g^{+}n\ket\otimes \overline{|j,g^{-}n\ket} &=& 
\sum_{m,m'} |j,m\ket \otimes \overline{|j,m'\ket} \int \mathrm{d}n\, t_{mj}^{j}(g^{+}n)\overline{t_{m'j}^{j}(g^{-}n)} \\
%&=& \frac{1}{d_{j}}\sum_{m,m'} |j,m\ket \otimes \overline{|j,m'\ket} \sum_{k} t_{mk}^{j}(g^{+})\overline{t_{m'k}^{j}(g^{-})}\\
&=&\frac{1}{\mathrm{d}_{j}}\sum_{m,m'} |j,m\ket \otimes\overline{|j,m'\ket}  t_{mm'}^{j}(g^{+}(g^{-})^{-1}).
\eeq
Using this result and some obvious changes of variables of integration, it is 
not hard to reduce (\ref{vertex-BC}) to the following, well-known
in the literature, expression of the vertex amplitude of the Barrett-Crane model:
\be\label{vertex-bc}
A^\sigma_{BC}[\{ j_{f} \}] = \int \prod_t dg_t \, \prod_f \mathrm{tr}_{j_f}((g_{t(f)})^{-1} g_{t'(f)}),
\ee
where $\{ j_f \}$ is the collection of 10 irreducible representations that
label the faces of $\sigma$, $t(f),t'(f)$ are the two tetrahedra of $\sigma$ that share 
the face $f$, and $\mathrm{tr}_j(g)$ is the character in $j$-th representation of an ${\rm SU}(2)$ 
element $g$. We thus see that the obviously wrong way of imposing the constraints
that ignores the fact that the face normals $n_{tf}$ in two neighbouring 4-simplices
should be the same produces the Barrett-Crane amplitude. The above derivation
both pinpoints what is wrong about the Barrett-Crane model, as well as 
suggests a way to remedy the problem.

One very important feature of the Barrett-Crane amplitude, which is also obvious from our analysis, 
is that it clearly corresponds to the gravitational and not the topological sector. This fact is in 
agreement with the analysis of Baez and Barrett \cite{Baez:1999tk} where it was shown, using very 
different arguments, that the Barrett-Crane way of implementing the constraints cannot lead to 
a quantisation of the topological sector.

It is illuminating to see the impossibility of having the topological sector ``Barrett-Crane'' model 
via our technology. Indeed, in an analog of the Barrett-Crane model for the topological sector, the amplitude
should be obtained by implementing the topological constraints (\ref{top1}) and therefore using the coherent states 
$|j,n_{tf}\ket\otimes |j,n_{tf}\ket$ instead of $|j,n_{tf}\ket\otimes \overline{|j,n_{tf}\ket}$.
However the integral 
\beq
\int \mathrm{d}n |j,g^{+}n\ket\otimes {|j,g^{-}n\ket} &=& 
\sum_{m,m'} |j,m\ket \otimes {|j,m'\ket} \int \mathrm{d}n\, t_{mj}^{j}(g^{+}n){t_{m'j}^{j}(g^{-}n)} \\
 &=& 
\sum_{m,m'} |j,m\ket \otimes {|j,m'\ket} \int \mathrm{d}n\, t_{mj}^{j}(g^{+}n)\overline{t_{m' j}^{j}(\epsilon g^{-}n\epsilon^{-1})} \\
%&=& \frac{1}{\mathrm{d}_{j}}\sum_{m,m'} |j,m\ket \otimes \overline{|j,m'\ket} \sum_{k} t_{mk}^{j}(g^{+})\overline{t_{m'k}^{j}(g^{-})}\\
&=&\frac{1}{\mathrm{d}_{j}}\sum_{m,m'} |j,m\ket \otimes\overline{|j,m'\ket}  t_{m m'}^{j}(g^{+}(\epsilon g^{-})^{-1}) t_{jj}(\epsilon)= 0
\eeq
is vanishing since $t_{jj}(\epsilon)=0$. Thus, one clearly sees that there is no way to impose the 
constraints of the topological sector \`a la Barrett and Crane on each vertex amplitude separately.

We have not discussed the closure constraints so far. However, these constraints are
automatically imposed by the integrals over the ${\rm SO}(4)$ holonomy group
elements $g_{\sigma t}$. Indeed, these integrals ``average'' the sum of all
four bivectors corresponding to a tetrahedron over all directions. Thus, only
the configurations that add up to zero survive the integration.

The above derivation has made it clear that, 
despite the important feature of the Barrett-Crane model as corresponding to the ``right'' 
gravitational sector of the theory, it is, however, plagued by a deficiency. Indeed,
as we have seen, in order to construct the Barrett-Crane model one has to integrate 
separately over the $n_{tf}^{\sigma}$ variables for each 4-simplex. But since the variables 
$n_{tf}^{\sigma}$ describe the bivectors corresponding to the faces of a tetrahedron $t$, 
the Barrett-Crane model describes ``geometry'' in which a tetrahedron $t$ viewed from one simplex
has a completely different ``shape'' as when viewed from a different 4-simplex. This is clearly not
a geometrically adequate way of implementing the constraints.  A related comment is that the Barrett-Crane vertex 
amplitude does not depend on the normals $n_{tf}$, which means that some of the geometric information about 
the geometry of the triangulated manifold is lost. It is possible that the fact 
\cite{Baez:2002rx}, \cite{Barrett:2002ur}, \cite{Freidel:2002mj}
that this vertex amplitude is dominated by degenerate tetrahedra is precisely due to this 
loss of geometrical information.

\subsection{The topological sector and the model of \cite{Engle:2007uq}}

From the analysis performed in the previous subsection it became clear that the main deficiency of the 
Barrett-Crane model is easy to resolve. The idea for how to do it is a simple extension of the set
of ideas developed in \cite{Freidel:1998pt}. Namely, the continuum partition function 
of BF theory is expressed as an integral over the connection field $A$ and the 2-form field $B$. 
In the previous section we saw that at the discrete level the BF theory partition function 
can be  expressed as a double integral over the connection (holonomies) and face variables ${\bf X}_{tf}$, see
(\ref{ZBF}), where ${\bf X}_{tf}$ are described by the group elements ${\bf n}_{tf}$. Similarly 
to the continuum theory,  one should impose the simplicity constraint at the 
level of the measure of integration over ${\bf X}_{tf}$. Importantly, since each tetrahedron  $t$ 
is shared by two 4-simplices $\sigma,\sigma'$ the same variable ${\bf X}_{tf}$ does appear in 
two vertex amplitudes. Imposing the constraints at the level of the integration measure over ${\bf X}_{tf}$ 
means that while we ask these bivectors to be simple we also impose the condition that 
${\bf X}_{tf}^{\sigma}= {\bf X}_{tf}^{\sigma'}$. Geometrically this corresponds to the condition that 
the geometry of the tetrahedron $t$  viewed  from the point of view of 4-simplex $\sigma$ agrees with the 
geometry viewed from $\sigma'$. 

We start by applying the above considerations to the topological sector of the theory. The
constraints we have to impose are: (i) instead of summing over all ${\rm SO}(4)$ representations 
${\bf j}_{f}$ we sum only over the simple representations ${\bf j}_{f}=(j_{f},j_{f})$ and (ii)
instead of integrating over all group elements ${\bf n}_{tf} $ we integrate 
over the normals of the form ${\bf n}_{tf}=(n_{tf} h_{\phi_{tf}},u_{t}  n_{tf} h_{\phi_{tf}}^{-1} )$.
After some simple changes of variables the integration the integrals over $u_{t}$ and $h_{\phi_{tf}}$ can 
be dropped and we are left with the following amplitude:
\beq\label{Ztop}
\tilde{Z}^{Top} &=& \sum_{{ j}_{f}} \prod_{f}\mathrm{d}_{{ j}_{f}}^{2}
\int \prod_{(t,f)}\mathrm{d}_{{ j}_{f}}^{2} \mathrm{d}{ n}_{t f}
\int \prod_{(t,\sigma)}\mathrm{d}{ g}_{\sigma t}^{+}\mathrm{d}{ g}_{\sigma t}^{-}
 \times\nonumber\\
 & &  \prod_{(\sigma,f) }\bra { j}_{f}, { n}_{tf}|  ({ g}_{ \sigma t}^{+})^{-1}{\ g}_{\sigma  t'}^{+}| { j}_{f}, { n}_{t'f}\ket
 \bra { j}_{f}, { n}_{tf}|  ({ g}_{ \sigma t}^{-})^{-1}{\ g}_{\sigma  t'}^{-}| { j}_{f}, { n}_{t'f}\ket
 \eeq
where all the integrals are over $SU(2)$ elements.

As in the case of the Barrett-Crane model it might seem that only two out of three constraints -- namely, simplicity
and cross-simplicity, but not the closure were imposed. However, as before, the integration over 
variables $g_{\sigma t}$ does impose the closure constraints since 
\be
\sum_{f\subset t} \hat{X}^{f}_{\pm}\cdot \left(\otimes_{f} \int {\mathrm d}g  |j_{f}, g h_{f}\ket \right) = 0.
\ee

It is clear from the expression above that the essential 
difference between this model and the model for BF theory is in the form 
of the intertwiner which is inserted for each pair $ft$. Indeed, instead of inserting the 
decomposition of the identity of the BF model:
\be
1_{j^{+}}\otimes 1_{j^{-}}=\mathrm{d}_{j^-} \mathrm{d}_{j^+} 
\int dn^- dn^+\, |j^-, n^-\ket\otimes |j^+,n^+\ket \bra j^-,n^- |\otimes \bra j^+,n^+|,
\ee
the model we are discussing is obtained by inserting the simple intertwiner 
\be
T_{j}\equiv \mathrm{d}_{j}^{2} 
\int \mathrm{d}n\, |j, n\ket\otimes |j,n\ket \bra j,n |\otimes \bra j,n|.
\ee

One can  compute this intertwiner by recalling the standard tensor product
isomorphism $V_{j}\otimes V_{j} =\oplus_{l=0}^{2j} V_{k}$, which is realised in terms of the group 
invariant maps. Thus, there exist invariant maps 
\be C^{j^{+}j^{-}k}: V^{j^{+}}\otimes V^{j^{-}}\to V^{k},\quad \overline{C^{j^{+}j^{-}k}}: V^{k} \to V^{j^{+}}\otimes V^{j^{-}}.\ee
These maps are unique up to normalisation, we chose the normalisation such that 
\be 
\sum_{k=|j^{+}-j^{-}|}^{j^{+}+j^{-}} \mathrm{d}_{k}\, \overline{C^{j^{+}j^{-}k}} C^{j^{+}j^{-}k}= 1_{j^{+}}\otimes 1_{j^{-}},\quad
 C^{j^{+}j^{-}k} \overline{C^{j^{+}j^{-}k'}} = \frac{\delta_{k,k'}}{\mathrm{d}_{k}} 1_{k}.
 \ee
The matrix elements of these intertwiners are the Clebsch-Gordan coefficients (3j symbols) 
and their complex conjugates
\be
C^{j^{+}j^{-}k}_{m^{+}m^{-}m} \equiv \bra k,m|C^{j^{+}j^{-}k}\left(|j, m^{+}\ket\otimes |j,m^{-}\ket \right),
\quad 
\overline{C^{jjk}_{m^{+}m^{-}m}} \equiv \bra j,m^{+} |\otimes \bra j,m^{-}|\overline{C^{j^{+}j^{-}k}}|k,m\ket .
\ee

It is now not hard to show that 
\be\label{2j}
|j,n\ket\otimes |j,n\ket =  {\sqrt{\mathrm{d}_{2j}}}\, \overline{C^{j}}  ( |2j, n\ket) ,
\ee
where we have denoted $C^{j}\equiv C^{jj2j}$. The fact (\ref{2j}) is of importance, so it
is worth seeing how this comes about in detail. Let us recall the definition
(\ref{def}) of the coherent states in terms of the matrix elements. Using this we
can write:
\be
|j,n\ket\otimes |j,n\ket = \sum_{m^-,m^+}  |j,m^- \ket\otimes |j,m^+\ket t^j_{m^-j}(n) t^j_{m^+ j}(n).
\ee
We now use the formula (\ref{a2}) for the product of two matrix elements involving the Clebsch-Gordan
coefficients to re-write:
\be
|j,n\ket\otimes |j,n\ket = \sum_{m^-,m^+,m,m'} \sum_{l=0}^{2j} {\rm d}_l \, 
 |j,m^- \ket\otimes |j,m^+\ket \overline{C^{jj k}_{m^- m^+ m}} C^{jj k}_{jj m'} t^k_{mm'}(n).
\ee
However, it is easy to see that the last Clebsch-Gordan coefficient is only non-zero for $m'=j+j=2j$,
and that the only representation for which this is possible is $k=2j$. The Clebsch-Gordan
coefficient of relevance is computed in the appendix, with the result being:
\be\label{c2j}
C^{jj k}_{jj m'} = \frac{\delta^{k\,2j}\delta^{2j\,m'}}{\sqrt{{\rm d}_{2j}}}.
\ee
This implies that
\beq
|j,n\ket\otimes |j,n\ket &=& \sqrt{{\rm d}_{2j}} 
\sum_{m^-,m^+,m}  |j,m^- \ket\otimes |j,m^+\ket \overline{C^{j}_{m^- m^+ m}} t^{2j}_{m 2j}(n)
\\ \nonumber
&=& \sqrt{{\rm d}_{2j}}  \sum_m \overline{C^{j}}( |2j, m\ket)  t^{2j}_{m 2j}(n)=  \sqrt{{\rm d}_{2j}} \,\overline{C^{j}}(|2j, n\ket),
\eeq
which proves (\ref{2j}).
We can now easily perform the integral
\be
T_{j} = \mathrm{d}_{j}^{2} \mathrm{d}_{2j} \int \mathrm{d}n \,\overline{C^{j}} |2j,n\ket \bra 2j,n | C^{j} =
{\mathrm{d}_{j}^{2}}\, \bar{C}^{j}\circ  C^{j}.
\ee
Note that this intertwiner  is proportional to a projector 
$T_{j}T_{i}= \delta_{i,j} (\mathrm{d}_{j}^{2}/\mathrm{d}_{2j}) T_{j}.$ 

This intertwiner is (up to normalisation) the one introduced 
by Engle, Pereira Rovelli in \cite{Engle:2007uq}. This shows that the model we obtain after 
integrations over the group elements $g_{\sigma t}^{\pm}$ is (up to a different edge normalisation) 
the model of \cite{Engle:2007uq}. However, it is clear from our analysis that this model gives a quantisation 
of the topological, not the gravitational sector. Thus, despite providing an important step in
the right direction, the model of \cite{Engle:2007uq} is unlikely to give an interesting
model of 4-dimensional quantum gravity.

For completeness of the exposition, let us show explicitly how one recovers the model 
of \cite{Engle:2007uq} as it is presented in this reference. First, let us introduce the 
following intertwiner:
\be
Y_{i}(j_{1},\cdots,j_{4})\equiv  
\sum_{m} C^{j_{1}j_{2}i}|i,m\ket \otimes {C}^{j_{3}j_{4}i}\overline{|i,m\ket}   
: V_{j_{1}}\otimes \cdots \otimes V_{j_{4}} \longrightarrow \mathbb{C}.
\ee
This intertwiner appears in the result of evaluation of the following group integral
\be\label{intg}
\int \mathrm{d}g \, \,t^{j_{1}}(g) \otimes \cdots \otimes t^{j_{4}}(g) = 
\sum_{i} \mathrm{d}_{i} Y_{i}(j_{1},\cdots,j_{4}) \bar{Y}_{i}(j_{1},\cdots,j_{4}).
\ee

The problem of computation of the partition function (\ref{Ztop}) of the topological spin foam 
model reduces to that of evaluation of the integral 
\be
I= \int \mathrm{d}g^{+}\mathrm{d}g^{-} \otimes_{i=1}^{4}   
\left(  \left( t^{j_{i}}(g^{+})\otimes  t^{j_{i}}(g^{-})\right) \overline{C^{j_{i}}}|2j_{i},n_{i}\ket \right).
\ee
This can be done by a repeated use of (\ref{intg}). One finds: 
\be 
I= \sum_{k,i^{+},i^{-}} \mathrm{d}_{i^{+}}\mathrm{d}_{i^{-}} 
\overline{Y_{i^{+}}}(j_{i}) \otimes \overline{Y_{i^{-}}}(j_{i}) f_{i^{+}i^{-}}^{k}(j_{i}) \, 
\left(\mathrm{d}_{k}Y_{k}(2j_{i})\otimes_{i=1}^{4}  |2j_{i},n_{i}\ket \right),
\ee
where 
\be
f_{i^{+}i^{-}}^{k}(j_{i}) \equiv \bra Y_{i^{+}}(j_{i}) 
\otimes Y_{i^{-}}(j_{i})|\otimes_{i=1}^{4}\overline{C^{j_{i}}}| \overline{Y_{k}}(2j_{i})\ket
\ee
are precisely the coefficients introduced in \cite{Engle:2007uq}. The vertex 
amplitude is then given by:
\be
A^{Top}(j_{f},k_{t}) = \sum_{i^{+}_{t},i^{-}_{t}} 
\mathrm{15j}_{SO(4)}(j_{f},j_{f},i^{+}_{t},i^{-}_{t}) \prod_{t} 
\mathrm{d}_{i^{+}_{t}}\mathrm{d}_{i^{-}_{t}} f_{i^{+}_{t} i^{-}_{t}}^{k_{t}}.
\ee
The dimension factors we obtain are not all included in  \cite{Engle:2007uq}.
Apart from that, the amplitude is exactly the one proposed in this reference.
The final model is obtained by summing over $k_{e}$ and $j_{f}$ with the measures
$\sum_{k}\mathrm{d}_{k}$ and $ \sum_{j}{\mathrm{d}_{j}^{2}}$ and by associating to each 
strand of the dual edge (equivalently each pair $tf$) a factor of ${\mathrm{d}_{j}^{2}}$.
This is the model described in \cite{Engle:2007uq} with a slightly different choice of 
the edge amplitudes. Note that the precise form of the edge amplitude depends on how the 
simplicity constraints are imposed. 

The model of \cite{Engle:2007uq} thus solves the main problem of the Barrett-Crane model: 
it does solder the geometry of one tetrahedron to the geometry of the same tetrahedron viewed from 
another 4-simplex. That in itself is a great success.  The use of the coherent states renders the 
construction leading to this model rather transparent in that it recovers it 
from simple and well-motivated operations on the partition function
of the BF theory. However, as we have already emphasized above, the model described 
gives a quantisation of the {\it topological} sector of the constrained BF theory.

\subsection{The gravitational sector: a new model}

The two models described above, namely that of Barrett-Crane and that of Engle et al each
has its own merit, but, as was clear from our analysis, both fail to provide a satisfactory 
model for 4d quantum gravity. Our analysis however also suggests a way to 
merge together the advantages of the Barret-Crane model 
(as corresponding to the gravitational sector) 
with the advantages of Engle et al. model (soldering the neighboring geometry).
This leads to a new spin foam model free of the described drawbacks.

The idea is essentially the same as the one already implemented in the case of the topological
sector. Thus, we impose the following two constraints on the partition function of the BF theory:
(i) only the simple representations will be included in the state sum and (ii) 
instead of integrating over all ${\rm SO}(4)$ group elements ${\bf n}_{tf} $ we integrate only over the one 
having the form ${\bf n}_{tf}=(n_{tf} h_{\phi_{tf}},u_{t}  n_{tf} \epsilon h_{\phi_{tf}}^{-1} )$.
As we have seen above the effect of the group element $\epsilon$ is to select the coherent states 
$|j,n\ket\otimes \overline{|j,n\ket}$ instead of $|j,n\ket\otimes |j,n\ket$.
After some simple changes of variables of integration the integrals over $u_{t}$ and $h_{\phi_{tf}}$ 
are again dropped and we are left with the following partition function:
\beq\label{Zgrav}
\tilde{Z}^{Grav} &=& \sum_{{ j}_{f}} \prod_{f}\mathrm{d}_{{ j}_{f}}^{2}
\int \prod_{(t,f)}\mathrm{d}_{{ j}_{f}}^{2} \mathrm{d}{ n}_{t f}
\int \prod_{(t,\sigma)}\mathrm{d}{ g}_{\sigma t}^{+}\mathrm{d}{ g}_{\sigma t}^{-}
 \times\nonumber\\
 & &  \prod_{(\sigma,f) }\bra { j}_{f}, { n}_{tf}|  ({ g}_{ \sigma t}^{+})^{-1}{\ g}_{\sigma  t'}^{+}| { j}_{f}, { n}_{t'f}\ket
 \overline{\bra { j}_{f}, { n}_{tf}|  ({ g}_{ \sigma t}^{-})^{-1}{\ g}_{\sigma  t'}^{-}| { j}_{f}, { n}_{t'f}\ket},
 \eeq
where all the integrals are over $SU(2)$ elements.

We therefore see that imposing the simplicity and cross-simplicity constraints 
as appropriate for the gravitational sector means that the following intertwiner
is inserted into each ``strand'' (corresponding to a face $f$) of each dual edge
(corresponding to a tetrahedron $t$) 
\be\label{g-j}
G_{j}\equiv \mathrm{d}_{j}^{2} 
\int \mathrm{d}n\, |j, n\ket\otimes \overline{|j,n\ket} \bra j,n |\otimes \overline{\bra j,n|}.
\ee
Using arguments similar to those in the previous section one can show that 
\be\label{tensgrav}
|j, n\ket\otimes \overline{|j,n\ket} = \sum_{k=0}^{2j} 
\mathrm{d}_{k}\sqrt{C_{k}^{j}} \, \overline{C^{jjk}} \left(t^{k}(n) |k,0\ket\right),
\ee
where the Clebsch-Gordan coefficient
\be 
C_{k}^{j}=\left(C^{jjk}_{j-j 0}\right)^{2}=  \frac{(2j)!}{(2j-k)!} \frac{(2j)!}{(2j+k+1)!}
\ee
is computed in the appendix and $|k,0\ket$ denotes the $\SU(2)$ 
invariant (spherical) state in the representation $V_{k}$.

To prove (\ref{tensgrav}) we consider the tensor product 
\beq
|j,j\ket\otimes |j,-j\ket &=& \sum_{k} \mathrm{d}_{k} \,  
\overline{C^{jjk}}\circ C^{jjk} \left(|j,j\ket\otimes |j,-j\ket \right) \\
&=&  \sum_{k,m} \mathrm{d}_{k} \,  \overline{C^{jjk}} |k,m\ket  {C^{jjk}_{j-j m}} =  
\sum_{k} \mathrm{d}_{k} \,  \overline{C^{jjk}} |k,0\ket  {C^{jjk}_{j-j 0}},
\eeq
where in the first line we have used the decomposition of the identity in terms of the 
Clebsch-Gordan coefficients, and in the second line the fact that the relevant Clebsch-Gordan coefficient 
is zero unless $m=0$. This establishes the result (\ref{tensgrav}) for the case $n={\rm Id}$. The general case 
follows by considering the action on this state of the matrix $t^{j}(g)\otimes t^{j}(g)$ and using the intertwining 
property of the Clebsch-Gordan coefficients. Using  (\ref{tensgrav}) one obtains:
\be\label{Ggrav}
G_{j} = \mathrm{d}_{j}^{2} \sum_{k=0}^{2j} \mathrm{d}_{k} C_{k}^{j} \, \overline{C^{jjk}} \circ C^{jjk}.
\ee
We note that for $k=0$ this intertwiner is the product of the usual Barrett-Crane 
intertwiners. Moreover,  as the coefficients $C^{j}_{k}$ decrease with $k$ increasing 
\be
C^{j}_{0} = \frac1{\mathrm{d}_{j}},\quad  C^{j}_{k} = \frac{\mathrm{d}_{j}-k} {\mathrm{d}_{j}+k} C^{j}_{k-1},
\ee
the Barrett-Crane intertwiner is in a certain sense the dominant term in the sum (\ref{Ggrav}).

To finish with the presentation of the model we describe it explicitly in terms
of the 15j symbols. Similarly to what we have done in the previous subsection,
we integrate over the ${\rm SO}(4)$ holonomy group elements. The result is
most conveniently presented by introducing the set of coefficients:
\be
f_{i^{+}i^{-}}^{l}(j_{i},k_{i}) \equiv \bra Y_{i^{+}}(j_{i}) 
\otimes Y_{i^{-}}(j_{i})|\otimes_{i=1}^{4}\overline{C^{j_{i}j_{i}k_{i}} }| \overline{Y_{l}}(k_{i})\ket.
\ee
In terms of these the vertex amplitude is given by:
\be
A^{Grav}(j_{f},k_{tf}, l_{t}) = 
\sum_{i^{+}_{t},i^{-}_{t}} \mathrm{15j}_{SO(4)}(j_{f},j_{f},i^{+}_{t},i^{-}_{t}) 
\prod_{t} \mathrm{d}_{i^{+}_{t}}\mathrm{d}_{i^{-}_{t}} f_{i^{+}_{t} i^{-}_{t}}^{l_{t}}(j_{f},k_{tf}).
\ee
The full model is obtained by summing over $k_{t}$, $j_{f}$ and $k_{tf}$ with the measures
\be
\sum_{j_{f}} \prod_{f}{\mathrm{d}_{j_{f}}^{2}}, \quad  
\sum_{l_{t}}\prod_{t}\mathrm{d}_{l_{t}},\quad
\sum_{k_{tf}}\prod_{tf}{\mathrm{d}_{j_{f}}^{2}}C^{j_{f}}_{k_{tf}}.
\ee
Let us note that, similar to what happens in the case of the model \cite{Engle:2007uq}, 
the boundary spin networks of the model described are 4-valent.
The ${\rm SO}(3)$ spins $k_{tf}$ label the edges while  
$l_{t}$ label the vertices of the boundary ${\rm SO}(3)$ spin networks. 

\section{The Immirzi Parameter}
\label{sec:imm}

An important quantity that figures prominently in the loop approach to quantum
gravity is  the Immirzi parameter, customarily denoted by $\gamma$. 
The importance of this parameter is illustrated by the fact that it enters in the combination
$\hbar G \gamma$ into the expressions for the spectra of geometrical operators, 
see e.g. \cite{Ashtekar:2004eh}. At the classical level this parameter is present in 
the classical action, and, even though does not modify the equations of motion of 
the theory, it does modify its symplectic structure.
This parameter arises in our context as being present in   
the relation between the Lie algebra valued two-form field $B^{IJ}$ of the BF formulation
of gravity and the frame field one-forms $e^I$. Thus, we have, in the case of non-trivial 
$\gamma$, see \cite{Holst:1995pc}:
\be\label{imm-B}
B^{IJ}= \left(\frac12\epsilon^{IJ}{}_{KL} + \frac1{\gamma} 
\delta^{[I}_{K}\delta^{J]}_{L}\right) e^K\wedge e^L.
\ee
One of the puzzles associated with the Barret-Crane model has always been the fact that 
it was not possible to construct a spin foam model incorporating the Immirzi parameter, 
see \cite{Livine:2001jt} for a discussion and some early developments on this problem.

We would now like to describe how the approach we have developed in the present paper sheds new
light on this problem. As we have already seen above, at the discrete (simplicial) level 
an important role is played by the relation between the Lie-algebra-valued variables $X^{IJ}_f$ and
the ``area'' bivectors $A^{IJ}_f$, see (\ref{X-A}). Introduction of the Immirzi parameter 
changes this relation to:
\be \label{XA-g}
X_{f}^{IJ} =\tilde{A}_{f}^{IJ} + \frac1{\gamma} {A}_{f}^{IJ}
\Longleftrightarrow A^{IJ}= \frac{\gamma}{1-\gamma^{2}}(X^{IJ} - \gamma \tilde{X}^{IJ}),
\ee
where we have assumed that $\sigma=1$ for simplicity (Riemannian signature).
The original gravitational case (\ref{X-A}) can be recovered from this general
relation by sending $\gamma\to\infty$. 

We have already seen that the geometrically correct way
to impose the simplicity and cross-simplicity constraints is to demand that 
there is a vector $n_{I}$ such that all the area bivectors $A^{IJ}_f$ 
are orthogonal to $n^I$. In the case of a non-zero Immirzi parameter 
this leads to the requirement:
\be
n_I A^{IJ}=0  \Leftrightarrow n_{I}X^{IJ}_{f} - \gamma n_{I} \tilde{X}^{IJ}_{f}=0.
\ee
When written in terms of the self-dual decomposition this is rephrased
as the requirement that there exist 
$X$ in $\mathfrak{su}(2)$ and $u\in \SU(2)$ such that 
\be\label{xsd-g}
{\bf X}_{f} = \left( \left(1+\frac{1}{\gamma}\right)X_{f}, 
-\left(1-\frac{1}{\gamma}\right)uX_{f}u^{-1}\right).
\ee

It is now clear that there are two distinct cases to consider: $\gamma>1$ and
$\gamma<1$ (only the first of these cases was discussed in the original arxiv version
of this paper). For $\gamma>1$, the quantum version of the simplicity constraints amount to 
the following restriction on the irreducible representations of ${\rm SO}(4)$:  
\be \label{imm-1}
\gamma>1 \qquad \qquad j^{\pm}= \frac{\gamma \pm 1}{\gamma} j, 
\ee 
where $j^{\pm}, j$ are half  integers. It is clear that this condition can be satisfied only if 
$\gamma$ is rational, and  given by 
\be \label{gammacond-more}
\gamma = \frac{j^{+}+ j^{-}}{j^{+}- j^{-}}>1.
\ee
Moreover, in the case $\gamma>1$ the coherent states to be used when imposing
the constraints are the ones in their ``anti-parallel'' version 
$|j,g\ket\otimes \overline{|j,{g}\ket}$, as is clear from
the presence of the minus sign in front of the second entry in (\ref{xsd-g}).

For $\gamma<1$ we have:
\be \label{imm-2}
\gamma<1 \qquad \qquad j^{\pm}= \frac{1\pm \gamma}{\gamma} j.
\ee 
The corresponding rational value of $\gamma$ is:
\be\label{gammacond-less}
\gamma = \frac{j^{+}- j^{-}}{j^{+}+ j^{-}}<1.
\ee
It is clear that in this case the coherent states to be used are the
``parallel'' ones $|j,g\ket\otimes |j,g\ket$.

From (\ref{gammacond-more}), (\ref{gammacond-less}) we see that the Immirzi parameter is 
quantized. The case $\gamma=\infty$ is the one previously studied where $j^{+}=j^{-}$.
Now, for either of the cases (\ref{gammacond-more}) or (\ref{gammacond-less}) let us 
write the irreducible representations of ${\rm SO}(4)$ compatible with the simplicity 
condition as:
\be
j^{\gamma \pm} = \gamma^{\pm} j,
\ee
where $j\in \mathbb{Z}/2$ is a half integer and $\gamma^\pm$ are as
follows from (\ref{imm-1}) or (\ref{imm-2}).

The above considerations allow us to write a spin foam model that corresponds to a 
non-zero $\gamma$. There are two different models, depending on whether $\gamma$ is
greater or smaller than unity. As before, one needs to specify only the intertwiner 
associated with each pair $tf$. For the $\gamma<1$ case the
appropriate intertwiner is given by:
\beq
\gamma<1 \qquad 
T_{j}^{\gamma} &\equiv&\mathrm{d}_{j\gamma^{+}}\mathrm{d}_{j\gamma^{-}} \int \mathrm{d}n \,
|j\gamma^{+}  , n\ket\otimes {|j\gamma^{-},n\ket} \bra j\gamma^{+},n |\otimes {\bra j\gamma^{-},n|}\\
&=&\mathrm{d}_{j\gamma^{+}}\mathrm{d}_{j\gamma^{-}}\, \overline{C^{j\gamma^{+} j\gamma^{-}  
j(\gamma^{+}+\gamma^{-})}}C^{j\gamma^{+} j\gamma^{-} j(\gamma^{+}+\gamma^{-})}.
\eeq
In the $\gamma>1$ case the intertwiner is:
 \beq
\gamma>1 \qquad 
G_{j}^{\gamma} &\equiv&\mathrm{d}_{j\gamma^{+}}\mathrm{d}_{j\gamma^{-}} \int \mathrm{d}n \,
|\gamma^{+} j , n\ket\otimes \overline{|\gamma^{-}j,n\ket} \bra \gamma^{+}j,n |\otimes 
\overline{\bra \gamma^{-}j,n|}\\
&=&\mathrm{d}_{j\gamma^{+}}\mathrm{d}_{j\gamma^{-}}\, 
\sum_{k=j(\gamma^{+}-\gamma^{-})}^{j(\gamma^{+}+\gamma^{-})} 
\mathrm{d}_{k}\left|C^{j\gamma^{+} j\gamma^{-} k}_{j\gamma^{+} j\gamma^{-} j(\gamma^{+}-\gamma^{-})}
\right|^{2}  
\,\,\overline{C^{\gamma^{+}j, \gamma^{-}j, k}}C^{j\gamma^{+} j\gamma^{-} k}.
\eeq
The relevant Clebsch-Gordan coefficients are computed in the appendix.

One sees that both $T^{\gamma}$ and $G^{\gamma}$ do not contain the trivial representation in 
their decomposition. This gives one explanation of why it was impossible to find them using the 
Barret-Crane way of imposing the constraints.

One specially interesting case is that of $\gamma=1$, or, equivalently $(\gamma^{+}=1,\gamma^{-}=0)$. 
In this case one might naively expect to get a self-dual model. However, as it can be seen from 
(\ref{XA-g}), this limit is singular and the geometrical interpretation  of $X_{f}$ is lost. 
Thus, the self-dual formulation of gravity should not be expected to arise from this limit,
at least not in a naive way. We note that, interestingly, 
for this value of $\gamma$ the intertwiners of both the 
gravitational and topological sectors  simplify 
and are equal to the identity operator
\be
G_{j}^{1}= T_{j}^{1} =  1_{j}.
\ee

\section{A new model: Lorentzian signature}
\label{sec:lorentz}

An added benefit of our construction is that it is possible
to generalize the whole discussion to the Lorentzian case.

In case of the Lorentz group we have an isomorphism ${\rm SO}(1,3)\sim{\rm SL}(2,\C)$.
This isomorphism can be realized explicitly by considering the action of ${\rm SL}(2,\C)$
on Hermitian $2\times 2$ matrices: ${\rm SL}(2,\C)\ni g: x\to g x g^\dagger$, where
$g^\dagger=(g^*)^T$ is the Hermitian conjugation. This transformation
preserves the determinant of $x$, and is thus an orthogonal transformation. As in the
Euclidean case, there is the ``canonical'' $\SU(2)$ subgroup, which is just the
natural $\SU(2)$ subgroup inside ${\rm SL}(2,\C)$ and which fixes the vector in $M^{1,3}$
that is represented by the identity matrix $x$. The Lorentzian Barrett-Crane model is obtained by 
integrating over precisely the described diagonal ${\rm SU}(2)$ subgroup in each strand forming the
dual edge. As we will see, a more illuminating way to derive the Lorentzian Barrett-Crane model
is via the coherent states. 

The key question for us is which coherent states should one work with?
Coherent states for the Lorentz group have been considered in the literature 
\cite{Perelomov,Pol'shin:1999qe}, but the context of these references is not quite what we need. 
One possible notion is to start with an $\SU(2)$ invariant state $|\rho,0\ket$ and act on it with 
a group element to define $|\rho,g\ket$. Because the state $|\rho,0\ket$ is $\SU(2)$-invariant, 
this state is labelled by a point in $H^{3}$. However, this is analogous to considering 
the coherent states arising from $|j,0\ket$ instead of $|j,j\ket$ in the case of $\SU(2)$, which
is not the correct prescription, as the uncertainty is not minimized in this case.
As we shall see below, the correct coherent states turn out to be labelled by a point 
on the (projective) null cone. The
following subsections develop the corresponding techniques in some detail. 
Once these states are identified the construction of the Lorentzian gravitational model follows 
the same lines as in the Euclidean case.

\subsection{${\rm SL}(2,\mathbb{C})$ simple representations.}

Let us first review how the irreducible unitary representations of the Lorentz group
are constructed. According to the general representations theory of non-compact groups
developed in e.g. \cite{Gelfand}, irreducible representations are easiest to describe in the
space of functions on the so-called space of horospheres in the group, which is the 
homogeneous group manifold $N\backslash G$, where $N$ is the largest nilpotent subgroup of $G$. For
$G={\rm SL}(2,\C)$ the nilpotent subgroup can be chosen to be the group $N_+$ of upper
diagonal matrices with $1$ on the diagonal. 
Let us parametrize the space ${\cal B}:=N_{+}\backslash {\rm SL}(2,\C)$ by two complex 
numbers $z_1,z_2\in\C^*$, where $\C^*$ is the extended complex
plane. The parametrization is as follows:
\be
b(z,w) = \left( \begin{array}{cc} z_2^{-1} & 0 \\  z_1 & z_2 \end{array} \right).
\ee
The right action of the group $G$ on ${\cal B}$ translates then to the so-called
affine action:
\be
 b(z_1,z_2) \left( \begin{array}{cc} a & b \\ c & d \end{array} \right)  \sim b(az_1+cz_2,bz_1+dz_2),
\ee
where the equivalence is that of the left multiplication by an $N_+$ group element. The
simplicity of the above action of $G$ on ${\cal B}$ allows for an immediate
description of the irreducible representations: they are realized in the spaces of 
homogeneous functions on ${\cal B}$ of some degree:
\be
F(\lambda z_1, \lambda z_2) = \lambda^{\sigma} \bar{\lambda}^{\sigma'} F(z_1,z_2), \qquad \lambda, \sigma,\sigma'\in \C.
\ee
Another, often more convenient realization of these representations is in the
space of functions of one complex variable, which is possible in view
of the above homogeneity property. Thus, we introduce a new complex variable $z=z_1/z_2$
on which the group acts by fractional linear transformations. We then have:
\be
F(z_1,z_2)=z_2^\sigma \bar{z_2}^{\sigma'} f(z),\quad f(z)\equiv F(z,1).
\ee
The transformed function is, on the other hand
\be
T_g \circ F(z_1,z_2) = (bz_1+dz_2)^{\sigma} \overline{(bz_1+dz_2)}^{\sigma'} F\left(\frac{az_1+cz_2}{bz_1+dz_2},1\right) :=
z_2^\sigma \bar{z_2}^{\sigma'} T_g \circ F(z,1),
\ee
which defines the action in the space of functions of one variable:
\be
T^{\sigma,\sigma'}_g\circ f(z) = (bz+d)^\sigma \overline{(bz+d)}^{\sigma'} f\left(\frac{az+c}{bz+d}\right).
\ee
A detailed analysis of \cite{Gelfand} shows that the {\it unitary} irreducible representations
are parametrized by a pair $(s,\rho), s\in\Z, \rho\in\R_+$ and correspond to:
\be
\sigma={-s- i\rho} - 1, \sigma'= {s-i\rho} - 1.
\ee
The group action is therefore:
\be\label{unitary-sl2}
T^{(s,\rho)}_g \circ f(z) = |bz+d|^{-2(1+i\rho)} \left( \frac{bz+d}{|bz+d|} \right)^{2s}
f\left(\frac{az+c}{bz+d}\right),
\ee
and the Hermitian product simply
\be
\bra f|g\ket = \frac{1}{\pi}\int d^{2}z\, \bar{f}(z)g(z).
\ee
The series of unitary irreducible representations $(0,\rho)$ and $(s,0)$ are
often referred to as {\it simple}. Loosely speaking, the two 
series can be referred to as simple ``continuous'' and ``discrete'' correspondingly.
Note, however, that there is no discrete series of representations of ${\rm SL}(2,\C)$
of the type familiar from the context of the real group such as ${\rm SL}(2,\R)$.
The simple representations play a special role for they admit a nice geometric
interpretation in terms of simple bivectors. 
Namely, let us compute the two Casimirs. Denoting by $J^{AB}$ the 
Hermitian generators of $SO(3,1)$ these are given by
\beq
\frac12 J^{AB}J_{AB} =   s^{2} -\rho^{2}-1,\quad \frac12 J^{AB}\tilde{J}_{AB} = 2 s \rho.
\eeq
Therefore, we see that $s=0$ or $\rho=0$ solves the simplicity constraints and that $s=0$ is the sector where
$J$ is timelike hence  $\tilde{J}$ spacelike.

There exists another  convenient realization of representations $(s,\rho)$ in terms
of functions (for the case of $s=0$) or sections of some bundle (for
$s\not=0$) over the null cone or over the hyperboloid $H_3$. The case of representations $(0,\rho)$
is the simplest one and we will restrict our attention to it. 

From now on we will use the Dirac bra-ket notation and denote $f(z)\equiv \bra f |\rho, z\ket$.
As we will see the states $|\rho,z \ket$ will be our coherent states in the Lorentzian case. 
In order to study their properties as well as their relation with the Euclidean coherent states 
it is  convenient to employ two different realisations of the representations $(0,\rho)$.

The first realization is obtained from the fact that
${\rm SO}(3,1)$ acts by its geometrical action on the space of functions on the null cone.
The simple representation $(0,\rho)$ is obtained by 
considering the functions which are homogeneous of 
degree minus $1+i\rho$, that is $\phi(\lambda\xi)=|\lambda|^{-1-i\rho}\phi(\xi)$.
The intertwining operator between this realisation and the realization (\ref{unitary-sl2}) is such 
that the value of this function on the lightcone along its intersection
with the plane $x_0+x_3=1$ exactly matches $f(z)$ (this intersection is the complex plane $\C^*$,
so this condition makes sense). Namely given a  complex number we can construct a null 
vector represented as zero determinant Hermitian matrix
\be
%\xi(z_{1},z_{2}) = \left( \begin{array}{cc} |z_{2}|^{2}& z_{1}\bar{z}_{2} \\ \bar{z}_{1}z_{2} & |z_{1}|^2 \end{array} \right)
%=|z_{2}|^{2}\xi\left(\frac{z_{1}}{z_{2}}\right) ,\quad  
 \xi(z) =\left( \begin{array}{cc} 1& z \\ \bar{z} & |z|^2 \end{array} \right).
\ee
and the identification is simply $\phi_{f}(\xi(z))=f(z)$.
In terms of the Dirac notation this means that we introduce states $|\rho,\xi\ket$ which are such that
\be
|\rho,z\ket=|\rho,\xi(z)\ket, \quad |\rho,\lambda \xi \ket = |\lambda|^{-(1+i\rho)} |\rho,\xi \ket.
\ee

The intersection of the null cone with the the plane $x_{0}=1$ is a two sphere.
We can therefore parametrise the null vector up to a scale by a point $n(z)$ on the 2-sphere.
For instance $\xi(z)=(1+|z|^{2})(1, \vec{n}(z))$ where $\vec{n}(z)= n(z)\sigma_{3}n(z)^{-1}$ 
and $n(z)$ is represented by a unitary matrix
\be
n(z)=\frac{1}{\sqrt{1+|z|^{2}}} \left( \begin{array}{cc} 1& z \\ -\bar{z} & 1 \end{array} \right).
\ee
We therefore have a realization in terms of function on $S^{2}$ or states $|\rho,n\ket$, 
the relation with other realization is given in term of the coherent states by
\be
|\rho,n(z)\ket = {(1+|z|^{2})^{1+i\rho}}\,  |\rho,z\ket.
\ee

The second realization is in terms of  functions on $H^{3}$.
The representation $(0,\rho)$  can be obtained from functions on $H^{3}$ which satisfy the wave equation
\be \Box_{H^{3}}\phi(X)= -(\rho^{2}+1)\phi(X).
\label{waveq}\ee
The scalar product is normalised to be
\be
\bra f|g\ket= \int_{H^{3}} \frac{\mathrm{d} X}{(2\pi)^{2}} \bar{f}(X)g(X).
\ee

The intertwining operator between this and the previous realization is given as follows.
One can  use  function on the light cone as the boundary 
data for such solutions of the wave equation (\ref{waveq}).
It is important to remark that a solution of (\ref{waveq}) can be obtained from a solution of the 
equation $\Box_{\R^{1,3}} \phi(X)=0$ provided that   $\phi(X)$ is extended inside the light cone 
as an homogeneous function of degree $-(1+i\rho)$ {\it or } of degree $-(1- i\rho)$.
The fact that there is these two possibilities is an important subtle point. It is related to 
the fact that the representations $(0,\rho)$ and $(0,-\rho)$ are equivalent.
More explicitly, the relation between the two realizations is as follows. 
A point $x\in H_3$ is represented by the following Hermitian matrix:
\be
X= \left( \begin{array}{cc} x_0+x_3 & x_1+ix_2 \\  x_1-ix_2 & x_0-x_3 \end{array} \right)
\ee
of unit determinant. Let us now define the function:
\be
K_{\rho}(X,z) := \bra \rho, X| \rho, z\ket := \left( {\rm tr}(\xi(z)\tilde{X} ) \right)^{-(1+i\rho)}= (2 \xi(z)\cdot X)^{-(1+i\rho)},
\ee
where $\tilde{X}\equiv \epsilon X^{t}\epsilon^{-1}$, $\tilde{X}=X^{-1} {\mathrm det}(X)$ if $X$ is invertible and 
$\cdot$ denotes the scalar product on $\R^{1,3}$.
 It is  convenient to use the upper half-space model for $H_3$,
in which it is coordinatized by $(t,y), t\in\R_+, y\in\C$, where 
\be
X(t,y)=\left( \begin{array}{cc} t & t y \\ t\bar{y} & t|y|^{2}+t^{-1}  \end{array} \right)
\ee
and $ds^{2}= dt^{2}/t^{2} + t^{2} dy^{2}$.
Then this function is explicitly given by
\be
K_{\rho}(X,z)= \frac{ 1}{ ( t|y-z|^2 +t^{-1})^{(1+i\rho)}}.
\ee
One sees from the definition that 
$K_{-\rho}(X,z)=\overline{K_{\rho}(X,z)}$ and in the following we will denote $K_{\rho}(z,X):=\overline{K_{\rho}(X,z)}$.

This quantity plays a prominent role in the so-called AdS/CFT correspondence of string theory
and is referred to in that context as bulk-to-boundary propagator. 
Given a real function in the hyperboloid we can use this propagator to reconstruct 
the function from its asymptotic data at infinity. The relation  $f\to \phi^{f}$ between the
realization in the space of functions on $\C$ and on $H_3$ is the explicitly given by:
\be\label{Bb}
\phi^{f}(X) = \int \frac{d^2z}{\pi} \, \overline{K_{\rho}(X,z)} f(z)
=\int \frac{d^2z}{\pi} \,\bra f|\rho,z\ket K_{\rho}(z,X).
%\,\,\,{\mathrm{ or}}\quad |\rho, X\ket = \int \frac{d^2z}{\pi} \,|\rho, z\ket  \bra \rho, z| \rho, X\ket.
\ee
where we have written this expression in terms of the coherent states.

Starting from the expression (\ref{Bb}) we can compute 
the asymptotics of the field from which we recover 
the function on the null cone. A non-trivial computation that is most 
conveniently done using the Fourier modes
$f(z)=e^{i(\bar{p}z+ p\bar{z})}$ gives 
\be\label{asympt}
\phi^{f}(X(t,y))\, {\stackrel{\sim}{{}_{t\to \infty}}} \,
t^{-(1+i\rho)}\, \frac{f(y)}{-i\rho} + t^{-(1- i\rho)} \,\int \frac{d^2z}{\pi} \frac{{f(z)}}{|y-z|^{2(1-i\rho)}}.
\ee
Note that the operator 
\be
K_{\rho}(y,z):=\frac{-i\rho}{|y-z|^{2(1+i\rho)}}
\ee
intertwines the representation $(0,-\rho)$ with the representation $(0,\rho)$. Moreover 
this operator is the kernel corresponding to powers of the 2-dimensional Laplacian as can be easily seen 
by taking its 2 dimensional Fourier transform, namely
\be
\frac{\Gamma(1-i\rho)}{\Gamma(1+i\rho)}
 \left|\partial_{z}\partial_{\bar{z}}\right|^{i\rho}f(z) = \int \frac{d^2z}{\pi}\,  K_{\rho}(z,y) f(y).
\ee
Accordingly we have 
\be
\int \frac{d^2z}{\pi}\,  K_{\rho}(z,y) K_{-\rho}(y,w) =K_{0}(z,w)=\pi \delta^{2}(z-w)
\ee

In terms of the bulk-boundary propagator the asymptotics (\ref{asympt})  reads 
\beq\label{decomp}
K_{\rho}(X(t,x),y) & {\stackrel{\sim}{{}_{t\to \infty}}}& \frac{ t^{-(1+i\rho)}}{-i\rho} K_{\rho}(x,y) +\frac{ t^{-(1-i\rho)}}{i\rho} K_{0}(x,y)
\eeq
The advantage of the representation on the hyperboloid is the fact that 
the intertwining operator between $(0,\rho)$ and $(0,-\rho)$ is trivial, 
that is $|-\rho,X\ket = |\rho, X\ket$,
thus $K_{\rho}(X,z) = \bra -\rho, X| \rho, z\ket $.
The scalar product between two states on the hyperboloid can be 
easily computed from (\ref{decomp})
\be
K_{\rho}(X,Y)=2 \int \frac{\mathrm{d}^{2}z}{\pi}\, K_{\rho}(X,z) K_{-\rho}(Y,z) = 2 \frac{\sin \rho r}{\rho \sinh r}
\ee
where $r$ is the hyperbolic distance between $X$ and $Y$. 
Since it is real we have $K_{\rho}(X,Y)= K_{-\rho}(X,Y)$.
One can recover the bulk-to-boundary propagator from its asymptotics, 
which can be evaluated directly
\be
K_{\rho}(X,Y(t,y)) \sim_{t\to \infty} 
\frac{ t^{-(1+i\rho)}}{-i\rho} K_{\rho}(X,y) +\frac{ t^{-(1-i\rho)}}{i\rho} K_{-\rho}(X,y),
\ee
from which we deduce using (\ref{asympt})
\be
K_{-\rho}(X,z)= \int \frac{\mathrm{d}^{2}z}{\pi}\, K_{\rho}(X,\omega)K_{-\rho}(w,z).
\ee
The asymptotics and correspondence between $(0,\rho)$ and $(0,-\rho)$ therefore reads
\beq
|\rho, X(t,x) \ket \, &{\stackrel{\sim}{{}_{t\to \infty}}}& \, \frac{1}{-i\rho} |\rho,t \xi(x)\ket  + \frac{1}{i\rho}|-\rho, t \xi(x)\ket,\nonumber\\
|-\rho,z\ket &=& \int \frac{\mathrm{d}^{2}\omega}{\pi}\, K_{-\rho}(z,\omega)|\rho,\omega \ket.
\eeq

\subsection{Coherent states.}\label{Lcoherent}

Coherent states for the simple representations $(0,\rho)$ are easiest to describe
in the realization in the space of functions on the hyperbolic space $H_3$. Such
a state is parametrized by a point $z\in\C$ and is given simply by:
\be
\phi_{|\rho,z\ket}(X):= K_{\rho}(X,z).
\ee
Thus, the coherent states are essentially the states $|\rho, z\ket$ introduced above. 
From this definition it is now possible to compute the scalar product between different coherent states
\beq
\bra \rho', z|\rho, \omega\ket &=& \int_{H^{3}} \frac{\mathrm{d}X}{2\pi^{2}}\,\overline{\phi_{|\rho',z\ket}(X)}\phi_{|\rho,z\ket}(X)=
\int_{H^{3}}\frac{\mathrm{d}X}{(2\pi)^{2}} \,K_{-\rho'}(X,x)K_{\rho}(X,\omega)
%\\&=& \int_{H^{3}}\frac{\mathrm{d}X}{2\pi^{2}} \left(2\xi(z)\cdot X \right)^{-(1-i\rho')}\left(2X\cdot \xi(\omega) \right)^{-(1+i\rho)}.
\eeq
One can first easily compute this convolution for general points on $H^{3}$
\be
\int_{H^{3}}\frac{\mathrm{d}Y}{(2\pi)^{2}} \,K_{-\rho'}(Y,X)K_{\rho}(Y,Z)=
 \frac{\delta(\rho'+\rho)}{\rho^{2}}K_{\rho}(X,Z) +\frac{\delta(\rho'-\rho)}{\rho^{2}}K_{\rho}(X,Z).
 \ee
 Using the asymptotics of the previous section we can now compute the scalar product between two coherent states 
 \be
\bra \rho', z|\rho, \omega\ket =\frac{\delta(\rho'-\rho)}{\rho^{2}}K_{0}(z,w)+\frac{\delta(\rho'+\rho)}{\rho^{2}}K_{\rho}(z,w) 
\ee
which is the result we were looking for.

The following identity is not hard to prove:
\be
\int_{0}^{+\infty} {\rho^{2}} {\mathrm{d}}\rho \, K_{\rho}(X,Y)= (2\pi)^{2} \delta(X,Y),
\ee
where the $\delta$-function on the right-hand-side is that on $H^{3}$.
Thus, we get the following decomposition of the identity 
\be
1_{H^{3}}= \frac12 \int_{-\infty}^{+\infty} {\rho^{2}} {\mathrm{d}}\rho  \int \frac{\mathrm{d}^2z}{\pi}\, |\rho, z\ket \bra \rho, z|
\ee
Note that the unity operator on $H^{3}$ is an operator that commutes with the action of $SU(2)$, 
$u1_{H^{3}}u^{\dagger} = 1_{H^{3}}$ for $u\in SU(2)$. It can be represented as an averaging operator 
acting on functions on $SL(2,\mathbb{C})$ 
\be
 1_{H^{3}}: L^{2}(G)\to L^{2}(SL(2,\mathbb{C})/SU(2))=L^{2}(H^{3}),\quad 1_{H^{3}} F(g)=\int_{SU(2)} {\mathrm{d}}u\, F(gu).
\ee
In order to make the comparison with the Euclidean case, 
it is convenient to work in term of the representation on $S^{2}=SU(2)/U(1)$.
The following decomposition of the identity holds:
\be
1_\rho = \int \frac{\mathrm{d}^2z}{\pi}\, |\rho, z\ket \bra \rho, z|= \int \frac{\mathrm{d}^2z}{\pi(1+|z|^{2})^{2}}\, |\rho, n(z)\ket \bra \rho, n(z)|=
\int_{SU(2)} \mathrm{d}n\, |\rho, n\ket \bra \rho, n|
\ee
These states are the Lorentzian analogs of the Euclidean states 
we used to construct the gravitational spin foam model.
Using these states it is possible to get both the Barrett-Crane model as well as a
new Lorentzian model rather quickly.

\subsection{Lorentzian gravitational model.}

Let us see how one can obtain the vertex amplitude of the Barrett-Crane model
using the coherent states. As in the Euclidean case, the idea is to insert into the partition
function of the BF theory the identity operator on $H^{3}$ that can be expanded in terms of 
a complete set of simple coherent states.
This prescription projects onto the set of simple representations and identifies 
the simple bivectors labelled by $\rho,n$ associated to neighboring tetrahedra.
After this is done, the amplitude for a triangulation
separates into pieces - the vertex and edge amplitudes. The vertex amplitude
is parametrized by $5\times 4$ quantities $z_{tf}$ ($n_{tf}$ in the
Euclidean case). The amplitude is given by:
\be\label{lorentz-new}
L^\sigma_{new}[\{\rho_f\}, \{ n_{tf} \}]=\int_{{\rm SL}(2,\C)} \prod_t dh_t \prod_f 
\bra \rho_f, n_{t(f)f} |  (h_{t(f)})^{-1} h_{t'(f)} | \rho_f, n_{t'(f) f} \ket ,
\ee
where as before $t(f),t'(f)$ are the two tetrahedra that share the face $f$ and we have denoted 
$\bra\rho,z |\rho,w \ket = K_{\rho}(z,\omega)$. 
Thus, the prescription for the gravitational model
is to take these vertex amplitudes (\ref{lorentz-new}),
multiply them (possibly using additional edge amplitude, see below), and then
integrate over the variables describing the normals keeping these variables
in the neighbouring simplices {\it the same}. All this is in exact analogy
with what happens in the Euclidean case. 

Now, as in the Euclidean case, the Barrett-Crane model is obtained by integrating
over the ``normals'' $n_{tf}$ independently, forgetting about the fact that
there is a neighbouring simplex that one should solder the amplitude to. 
One obtain a model in which we have the characteristic Barrett-Crane  
decoupling of the vertex amplitudes.
All this is in exact analogy with what happens in the Euclidean case. 

The effect of integrations over the variables $n_{tf}$ is to produce the state:
\be
|\rho,0\ket := \int_{SU(2)} \mathrm{d}n\, 
|\rho, n\ket= \int \frac{\mathrm{d}z}{\pi}\, \frac{1}{(1+|z|^{2})^{1-i\rho}}|\rho,z\ket,
\ee
which is the $\SU(2)$ invariant state in the representation $(0,\rho)$. The
function $\bra \rho, X|\rho, 0\ket$ in $H_3$ is known as the spherical function. After the
integration, the vertex amplitude of the Barrett-Crane model becomes:
\be\label{bc-lor-vertex}
L^\sigma_{BC}[\{\rho_f\}] = \int_{{\rm SL}(2,\C)} \prod_t dh_t  
\prod_f \bra \rho_f, 0| (h_{t(f)})^{-1} h_{t'(f)} | \rho_f, 0 \ket .
\ee
The quantity
\be
\bra \rho, 0| h^{-1} h'| \rho, 0\ket
\ee
is actually a function of two points on $H_3$ (in view of $\SU(2)$ invariance
of the states $|\rho, 0\ket$). This function is the bulk-to-bulk
propagator, which has been constructed in the previous subsection.
The integrals in (\ref{bc-lor-vertex}) then reduce to a multiple integral
over the hyperbolic space of a product of bulk-to-bulk propagators - one
for each face of the 4-simplex in question. The quantity (\ref{bc-lor-vertex}) is thus the
well-known vertex amplitude that was first introduced in \cite{Barrett:1999qw}.

To finish the description of the model it remains to give the set of edge
amplitudes. One possibility is to set these amplitudes to be just the
$\delta$-functions, similar to what we have done in the Euclidean context, see
(\ref{edge-BF}). Another possibility is to consider
a non-trivial edge amplitude constructed in exactly the same way as the
vertex amplitude (this is also possible in the Euclidean context, but
we refrained from discussing it above). This edge amplitude is more natural from the perspective
of group field theory, and is arguably the correct way to allow for
a non-trivial holonomy between the 4-simplices. The amplitude is parametrized
by $2\times 4$ variables $z_f, z'_f$ and is given by:
\be\label{lor-edge-new}
L^t_{new}[\rho_{f_i},z_{f_i}, 
z'_{f_i}] = \int_{{\rm SL}(2,\C)} dh\,\prod_{i=1}^4
\bra \rho_{f_i}, z_{f_i} | h | \rho_{f_i}, z_{f_i} \ket .
\ee

The expressions (\ref{lorentz-new}), (\ref{lor-edge-new}) are formal, as all the above 
integrals over the Lorentz group are likely to diverge. Thus, they must be given
 sense via an appropriate regularization procedure, 
as is also the case for the Lorentzian Barrett-Crane model, see \cite{Barrett:1999qw}. It is possible
that the same regularization procedure \cite{Barrett:1999qw} may work for the new model.
Also the amplitude for the whole triangulation is obtained by multiplying the vertex and edge
amplitudes, and then integrating over the variables $z_{tf}$ using as the measure the
usual $d^2 z_{tf}$ measure on the complex plane. This operation may again
result in divergences that have to be taken care of. After this is done, one has to integrate
over the representation labels $\rho_f$, which is likely to result in additional divergences.
All this has to be analysed in detail before it can be claimed that the model introduced
above is well-defined. We leave this work to future publications. 

It would be
important to study in more details 
the derivation sketched  in  section \ref{Lcoherent} in which the simplicity
constraints are imposed by an integral over an appropriate subgroup of the Lorentz group. 
Such a generalization will give a clearer geometrical perspective on the
constructions we have introduced. It would
also be quite interesting to find a model that would give the Lorentzian
counterpart of the Euclidean model introduced by \cite{Engle:2007uq}. In analogy
to what happens in the Euclidean case it is to be expected that this other model
will correspond to the topological sector. It is clear that this model should be constructed 
using some other type of coherent states. Such
a construction would be quite important for it would give a much better
understanding in which sense the model introduced here corresponds to the
gravitational sector of the theory.
Again, we leave an analysis of these and related issues to future publications.

Finally, let us note that unlike in the Euclidean case, there is not one,
but two very distinct Lorentzian Barrett-Crane models. Indeed, in its
most simple form the Barrett-Crane model is obtained by integrating
over the diagonal ${\rm SU}(2)$ inserted in every strand forming the dual edge.
This selects from the sum over representations only the simple representations
of the type $(0,\rho)$. It has been argued in \cite{Rovelliperez} that one can introduce 
the simple discrete representation $(s,0)$ by integrating over the subgroup $SL(2,\mathbb{R})$.
Such a prescription can be adapted to our model by inserting instead of 
$1_{H^{3}}$ the identity operator 
$1_{dS^{3}}$ on de Sitter space. A detailed 
geometrical analysis of this other Lorentzian model is required.

\section{Group Field theory}
\label{sec:gft}

The group field theory (GFT) version of the new models can be obtained quite easily,
as our presentation above was very much in the GFT spirit.
Compared to the original group field theory setup for the Barret-Crane model the 
modifications are minor in that one only needs to make some obvious modifications of 
the kinetic and interaction terms. First, consider the following functional on $SO(4)$:
\be 
T(g)=   \sum_{j} \rd_{2j} tr_{V^{j}\otimes V^{j}} \left( g T_{j}\right)
\ee
where $g=(g^-,g^+) $ and $\rd_{2j} I_{j}$ is the projector 
\be
T_j := \int dn\, | j,n\ket\otimes|j,n\ket \bra j,n|\otimes\bra j,n|
\ee
that we encountered above. We define the propagator to be 
\be
K(g_{i}, \tilde{g}_{i}) \equiv \int_{SO(4)} du dv  
\prod_{i=1}^{4} T(u g_{i}  \tilde{g}_{i}^{-1} v^{-1}).
\ee

The topological spin foam model can then be obtained 
as the Feynman graph expansion of a group field theory 
with the following quadratic term in the action: 
\be
\int_{SO(4)} \phi(g_{i})K(g_{i}, \tilde{g}_{i})\bar{\phi}(g_{i}).
\ee
The interaction term is given by an  expression identical to the one usually used 
\cite{DePietri:1999bx, Freidel:2005qe}.

The gravitational spin foam model is constructed in the same way but starting with the intertwiner
\be
G_j := \int dn\, | j,n\ket\otimes\overline{|j,n\ket} \bra j,n|\otimes\overline{\bra j,n|}
\ee
instead of $T_{j}$.
The Lorentzian model group field theory is constructed
similarly.

\section{Alternative way to impose the constraints}
\label{geom}

In the previous sections we have heavily used the coherent state techniques in order
to constrain the discrete analog of the B field in a way that solders the geometry of the
neighboring tetrahedra. In this section we will shown that it is possible to construct the
same models from a slightly different, more geometrical point of view. The idea is to restrict
the integration over the connection in a geometrical manner.

Given a four simplex $\sigma$, a discrete  $SO(4)$ connection on $\sigma$ is a collection  
of group elements $\bg_{\sigma t}$ and $\bg^\sigma_{tf}$, where $\bg_{\sigma t}$ represents the holonomy 
of the spin connection along  the segment going from the center of $\sigma$ to the center of a 
tetrahedron $t$, and  $\bg^\sigma_{tf}$ represents the holonomy from the center 
of $t$ to the center of a face $f$. Our conventions are such that $\bg_{\sigma t}=\bg_{t\sigma }^{-1}$,
$\bg_{tf}^{\sigma}=(\bg_{ft}^{\sigma})^{-1}$. Given a simplicial complex $\Delta$,
a discrete $SO(4)$ connection on $\Delta$ is the choice of a discrete connection 
$\bg_{\sigma t}, \bg_{t f}^{\sigma}$  for each 4-simplex. A discrete 2-form field on $\Delta$ is 
an assignement of a Lie algebra element ${\bf X}_{f}$ to each face $f$ of $\Delta$.

Given such a discrete ${\rm SO}(4)$ connection and two form field we can write the discrete BF action as a
pairing between the variables $X_{f}$ and the holonomy (curvature) associated with each pair 
$(\sigma, f)$ (such a pair is called a ``wedge'' in the litterature). This action is
exactly analogous to (\ref{disaction}), except that now the sum is taken over the wedges,
not (dual) faces:
\be
S=\sum_{(\sigma,f)} \mathrm{tr}\left({\mathbf X}_{f}G_{f}^{\sigma}\right), \qquad 
G_{f}^{\sigma}\equiv  \bg_{ft}^{\sigma}\bg_{t\sigma } \bg_{\sigma t'}\bg_{t' f}^{\sigma},
\ee
where $t,t'$ are the two tetrahedra of $\sigma$ sharing the face $f$.
The quantisation of this theory is given by the following path integral:
\be\label{ZZint}
\sum_{\mathbf{j}_{f}} \prod_{f}{\mathrm{d}}_{{\mathbf{j}}_{f}}^{2}\int \prod_{f}\mathrm{d}{\bf n}_{f}  \int  \prod \rd{\bf g}_{t\sigma} 
\rd{\bf g}_{tf}    \prod_{(\sigma,f)}
\bra {\mathbf j}_{f},{\mathbf n}_{f}|G_{f}^{\sigma}|{\mathbf j}_{f},{\mathbf n}_{f}\ket.
\ee
The BF theory is obtained by restricting the discrete connection to be such that 
$\bg_{ft}^{\sigma}= \bg_{ft}^{\sigma'}$ for neighboring 4-simplices.
This way of writing the $BF$ theory amplitude is equivalent to what
we have done in the earlier sections, where we have assigned only
a single quantity $X_{f}$ (and hence $n_{f}$) per face. The main result of this section is that 
the new gravitational models considered above can be obtained from this integral 
by imposing constraints on the discrete connection alone.

In order to describe this restriction on the connection to be integrated over
we need to introduction an extra structure on $\Delta$ that we call  ``geometrical data''.
By definition ``geometrical data'' on  $\Delta$ is an assignment of a set 
$(e_{\sigma}^{A}, {\mathbf X}_{f}, {u}_{t}^{\sigma})$ of the orthonormal frame $e_{\sigma}^{A}$ 
at the center of $\sigma$, a set of bivectors ${\mathbf X}_{f}$ at the center of each face $f$ 
and a set of unit vectors $u_{t}^{\sigma}$ at the center of each tetrahedron $t$ inside $\sigma$.

Given  an orthormal frame $e_{\sigma}^{A},  A=0,1,2,3$ it is convenient to think of unit vectors $e^A$ as elements
of ${\rm SU}(2)\sim S^3$. The normalization condition is then $\mathrm{tr}(e^{A}(e^{B})^{-1})=2 \delta^{AB}$. 
The frame $e_\sigma^A$ should be thought of as ``sitting'' at the center of $\sigma$.
Now, given a frame at $\sigma$ and a discrete connection in $\sigma$ , one can parallel transport this frame
to the centers of tetrahedra $t$ of $\sigma$ as well as the centers of faces $f$ using the discrete connection. 
Let us introduce a special notation for this transported frame:
\be
e_{t}^{A}\equiv (\bg_{t\sigma} e_{\sigma})^{A}= g_{t\sigma}^{+}e_{\sigma}^{A}g_{\sigma t}^{-},\quad
e_{f}^{A}\equiv (\bg_{ft} e_{t})^{A}= g_{ft}^{+}e_{t}^{A}g_{t f}^{-},
\ee
where $g^\pm$ are ${\rm SU}(2)$ group elements representing an ${\rm SO}(4)$ one.

Let us now state a condition of compatibility between the ${\rm SO}(4)$ discrete connection and the geometrical data. 
Thus, recall that given a bivector ${\bf X}$ it is always possible to chose an orthonormal oriented 
frame $e^{A}$ such that ${\mathbf X}$ assumes the following simple form
(the normal form) ${\mathbf X} =  \alpha\, e^{1}\wedge e^{2}+ \beta\, e^{0}\wedge e^{3}$ 
where $\alpha, \beta$ are subject to the condition  $\alpha \geq|\beta|$ but otherwise arbitrary.
Now given a frame $e^{A}_{\sigma}$ and a discrete ${\rm SO}(4)$ connection we can construct from it 
the ``reference'' geometrical data by chosing for each tetrahedron the vector $e_{t}^{0}$ and a family of bivectors 
${\mathbf X}_{f}(\alpha, \beta):=  \alpha_{f}\, e^{1}_{f}\wedge e^{2}_{f} + \beta_{f}\,  e^{0}_{f}\wedge e^{3}_{f}$.
We say that a discrete connection $(\bg_{\sigma t}, \bg_{t f}^{\sigma})$ is {\it compatible} with 
the geometrical data $(e_{\sigma}^{A},{\mathbf X}_{f}, { u}_{t}^{\sigma})$ if 
the following conditions are satisfied: (i) $u_{t}^{\sigma} =e^{0}_{t}\equiv g_{t \sigma}e^{0}_{\sigma}$;
(ii) ${\mathbf g}_{tf}\cdot{\mathbf X}_{f}$ is equal to $ {\mathbf X}_{f}(\beta, \alpha)$ up to a rotation fixing 
$u_{t}$ and for a choice of parameters $(\alpha, \beta)$ satisfying $ \alpha \geq |\beta|$.
Note that these conditions are not constraints on bivectors $X_{f}$, but those on the frame fields and hence 
the connection. In particular, we do not assume that the  bivector fields are simple.
These conditions are quite natural. Indeed, the first one ensures that the chosen normal vector is transported consistently 
inside the 4-simplex. The second one ensures that the parallel transport inside tetrahedra preserves this normal vector. 
Note that the parallel transported face vector ${\mathbf g}_{tf}\cdot{\mathbf X}_{f}$ is exactly the face 
bivector denoted ${\bf X}_{tf}$ in the previous sections. The additional restriction on the parameters $(\alpha,\beta)$ 
resolves a remaining discrete ambiguity. We can chose a dual condition $\beta \geq|\alpha|$ which will result in 
a different model.

Our main claim in this section is that the new gravity model discussed earlier in this paper 
(corresponding to the gravitational sector of the theory) can be obtained from the integral 
(\ref{ZZint}) by restricting the discrete connection to be compatible (in the sense just defined) with 
the geometrical data $(e_{\sigma}^{A},{\mathbf X}_{f}, {u}_{t}^{\sigma})$ and integrating out $u_{t}^{\sigma}$. 
Because of gauge invariance, after such a restriction and integration the model is independent of a choice of the
frame $e^{A}_{\sigma}$.

{\bf Proof:} For definitness we fix the frame $e_{\sigma}$ to be the canonical one 
$e_{\sigma}^{0}=1, e_{\sigma}^{i}= \sigma^{i}$. The first condition the reads 
$e_{\sigma}^{0}= g_{\sigma t }^{+}u_{t}(g_{\sigma t }^{-})^{-1}$, and is solved by 
${\bf{g}}_{\sigma t } =( h_{\sigma t},h_{\sigma t} u_{t})$, where $(h_{t\sigma},h_{t\sigma})$ is 
an $\SU(2)$ element that fixes the vector $e^{0}_{\sigma}=1$.
To solve the second condition let us denote by ${\mathbf n}_{f} = (n^{+}_{f},n^{-}_{f})$ 
the ${\rm SO}(4)$ rotation which brings ${\mathbf X}_{f}$ into its normal form with respect to $e^{A}_{f}$:
${\mathbf n}_{f}^{-1}\cdot X_{f}= \alpha\, e^{0}_{f}\wedge e^{3}_{f} + \beta\, e^{1}_{f}\wedge e^{2}_{f}$ 
with $\alpha \geq|\beta|$. Applying ${\bf e}=(1,\epsilon)$ we can exchange the role of $\alpha$ and $\beta$, thus 
$X_{f} = {\mathbf n}_{f}{\mathbf e}^{-1}\cdot  X_{f}(\beta, \alpha)$.
The second condition is then solved by  $(g^{+}_{tf}n_{f}^{+},g_{tf}^{-}n_{f}^{-}) = 
(h_{tf}, h_{tf})(h_{\phi_{tf}},h_{-\phi_{tf}})(1,\epsilon)$
where $h_{tf}$ is an arbitary $\SU(2)$ element and $\phi_{tf}\in [0,2\pi]$.

This shows that an ${\rm SO}(4)$ connection compatible with the  
geometrical data $(u_{t},X_{f})$ is parametrized by two $\SU(2)$ group elements $h_{t\sigma}, h_{tf}$ and one ${\rm U}(1)$ 
element $h_{\phi_{tf}}$. The matrix element entering the integral (\ref{ZZint}) is then given by
\beq
\bra {\mathbf j}_{f},{\mathbf n}_{f}|G_{f}^{\sigma}|{\mathbf j}_{f},{\mathbf n}_{f}\ket &=&
\bra { j}_{f}^{+},{ j}_{f}^{+} |h_{ft}^{\sigma}h_{t\sigma}h_{\sigma t'} h_{t'f}^{\sigma} 
|{ j}_{f}^{+},{ j}_{f}^{+}\ket e^{i(j^{+}-j^{-})(\phi_{t'f}- \phi_{tf})}\\
&\times&\bra { j}_{f}^{-},-{ j}_{f}^{-} |h_{ft}^{\sigma}(u_{t}^{\sigma})^{-1}h_{t\sigma}^{\sigma}h_{\sigma t'} 
u_{t'}^{\sigma} h_{t'f} |{ j}_{f}^{-},-{ j}_{f}^{-}\ket.
\eeq
The gravitational model is then recovered by integrating over all these data with the condition that 
$h_{tf}^{\sigma}=h_{tf}^{\sigma'}$, which implies that
$X_{tf}^{\sigma}=X_{tf}^{\sigma'}$, where ${\mathbf g}_{tf}^{\sigma}\cdot {\mathbf X}_{f}$.
Note that the condition $\phi_{tf}^{\sigma}=\phi_{tf}^{\sigma'}$ is not required by the 
geometrical condition $X_{tf}^{\sigma}=X_{tf}^{\sigma'}$.
The integral over these ${\rm U}(1)$ group elements therefore imposes the simplicity condition
$j^{+}=j^{-}$, without having to put it by hand.
It is relevant to note  that  this model does not require that $u_{t}^{\sigma}=u_{t}^{\sigma'}$.
If one relaxes the condition $X_{tf}^{\sigma}=X_{tf}^{\sigma'}$ one obtains the Barrett-Crane model, as we 
have seen in section \ref{sec:euclid}.

It is interesting to note that the above analysis suggests that one can impose an even more stringent 
geometrical constraint: $u_{t}^{\sigma}=u_{t}^{\sigma'}$, which insures that not only the internal but also the 
external geometry (the normals to the tetrahedra) agree. This gives an even more constrained model that would be 
interesting to study. However, from the canonical point of view the external and internal geometries are 
``canonically conjugate'', so it is probably not very natural to impose constraints on both data. 
However, more analysis is necessary  before a final conclusion on this can be reached.
Note also that this fully constrained version amounts to relaxing the way the closure constraints are
implemented and imposing them at the operator level instead of the level of the states.
Indeed, for each tetrahedron there is a double integration $\int du_{t} du'_{t}  u_{t} I u_{t'}^{-1}$, 
where $I$ is a certain operator.
This double integral allows to split $I$ as a product of intertwinners, which means that the closure constraints 
are imposed strongly. If one restricts $u_{t}=u_{t}'$, one gets only one integral left, namely
$\int du_{t} u_{t} I u_{t}^{-1}$, and the closure constraint is imposed weakly.
Note finally that we could also decide to constrain $u_{t}$ and not $X_{f}$, 
the physical meaning of which would be interesting to unravel.

To summarise, there are three different ways to impose the constraints on the geometrical data:
(i) we can fully constrain the data by imposing $u_{t}^{\sigma}=u_{t}^{\sigma'}$ and  
$X_{tf}^{\sigma}=X_{tf}^{\sigma'}$; (ii) partially constrain the data by imposing either 
$u_{t}^{\sigma}=u_{t}^{\sigma'}$ or  $X_{tf}^{\sigma}=X_{tf}^{\sigma'}$, which is the way of imposing the 
constraints we have studied in this paper; (iii) impose no constraints, which leads to the Barrett-Crane model.

\section{Conclusions}

Let us summarize the results of this paper. Using Plebanski (or BF) formulation
of gravity as the starting point we have described how the simplicity constraints
that convert the topological BF theory into a gravitational one can be imposed
using the coherent state techniques. We have seen how the Barrett-Crane model
\cite{Barrett:1997gw} gives a quantization of the correct gravitational sector of the theory but
decouples the neighbouring simplex amplitudes in an unnatural way. We have seen
how this decoupling or ultra-locality problem is resolved in the model of
Engle et al \cite{Engle:2007uq} and how this model corresponds to the topological
sector of the theory instead of the gravitational one. Finally, we have constructed
a new model, both in the Euclidean and in the Lorentzian settings. Both models
do not suffer from the ulta-locality problem characteristic of the Barrett-Crane models.
The Euclidean model we obtained can be seen to correspond to the ``correct'' gravitational
sector of the theory. In the case of the Lorentzian model further analysis is necessary
to establish whether it describes the gravitational or the topological sector.
However, as the Lorentzian model we constructed is the ``delocalized'' Barrett-Crane
model, this strongly suggests that the model corresponds to the gravitational sector.

Importantly, we have described how the Immirzi parameter $\gamma$, which at the level 
of the continuous formulation of the theory appears as a parameter in front
of the ``topological'' term in the action can be incorporated into 
the spin foam framework. Importantly, the cases $\gamma>1$ and $\gamma<1$ correspond to
models of a rather different nature. We have seen why it was impossible to 
incorporate $\gamma$ into the framework of spin foams via the Barrett-Crane way of imposing
the constraints. We have also seen that in the case of Riemannian signature models the Immirzi 
parameter turns out to be quantized, see (\ref{gammacond-more}) and (\ref{gammacond-less}) for
the corresponding expressions.

In spite of a number of interesting advances made in the present work, a
great deal remains to be done. Thus, it is necessary 
to repeat the analysis of the semi-classical limit of the vertex amplitudes
for the new models and see whether the amplitudes are dominated by 
non-degenerate geometrical configurations. It is necessary to develop
the Lorentzian signature model(s) in more detail, and, in particular,
identify the model that corresponds to the 
topological sector. After this is done it is important to study the relation
with the picture of quantum geometry that arises from loop quantum gravity.
It is not unlikely that the projected spin networks of \cite{Alexandrov:2002br}
will be an important part of this relation, see also a more recent work 
\cite{Alexandrov:2008da} for some steps in this direction. It is also interesting to
develop in more details the group field theory perspective on the new models.

At a higher level of complexity, it is essential to continue with the
program of the path integral quantization. Thus, in general in the path integral 
one has to impose not only the simplicity constraints, but also other second-class 
constraints, as well as gauge-fix the first class constraints (even though
there may be some simplifications if one is only interested in the partition
function, the case we have explored in the present work). So far, all these 
other steps have been missing from the framework of spin foam models of 
four-dimensional general relativity. It is unlikely
that further significant progress can be made without tackling these
issues. Another important point that has so far been missing from the
spin foam framework is understanding in some explicit fashion how to discretize
the Lagrange multipliers that impose the simplicity constraints. Indeed, 
recall that (some of) the Lagrange multipliers in question receive (on-shell) the important 
interpretation of being the Weyl part of the Riemann curvature tensor.
This is exactly the part of the curvature that describes the
local degrees of freedom of the gravitational field. Hence, it
would be very important to see explicitly how the spin foam
models of the type described in this work encode this Weyl
part of the curvature. 

In spite of the spin foam quantization programme being far from
complete, some preliminary physical computations of the type
\cite{Alesci:2007tx} are possible. It remains to be seen what the physics described by the new
model(s) is, but the absence of ultra-locality, which seemed to be
at the root of most of the problems suffered by Barrett-Crane
model(s), suggests that a new exciting period of the development
of the subject of spin foam models is opening.

{\bf Acknowledgement:} We would like to thank D. Oriti for collaboration on the very initial 
stages of this project and E. Livine, S. Speziale for discussions of their work.

\section*{Appendix: Some necessary formulae}

The following two standard formulae are often used in the present paper:
\be\label{a1}
\int_G dg t^{j}_{mn}(g) \overline{t^{j'}_{m'n'}(g)} = \frac{\delta^{jj'}}{\rd_j} \delta^{mm'}\delta^{nn'}.
\ee
\be\label{a2}
t^{j_1}_{m_1n_1}(g) t^{j_2}_{m_2n_2}(g) = \sum_{j=|j_1-j_2|}^{j_1+j_2} \sum_{mn} 
\rd_j \overline{C^{j_1j_2 j}_{m_1 m_2 m}} 
C^{j_1 j_2 j}_{n_1 n_2 n} t^j_{mn}(g).
\ee
The last formula, in particular, fixes our choice of the normalization of the Clebsch-Gordan coefficients.

\section*{Appendix: Clebsch-Gordan coefficients}\label{clebsch}

The general formula for the Clebsch-Gordan coefficient of ${\rm SU}(2)$ (in the normalization that
the $\theta$-symbol is equal to one) is given by:
\be\label{cg-gen}
C^{l_1 l_2 l}_{j k m} = 
(-1)^{l_1-j} \frac{ [\vec{l},\vec{j}] (l_1+l_2-l)!}
{\Delta(\vec{l})  } \\ \nonumber
\sum_r \frac{(-1)^r}{r!} \frac{(l_1+j+r)!(l_2+l-j-r)!}
{(l-m-r)!(l_1-j-r)!(l_2-l+j+r)!},
\ee
where the sum is taken over all values of $r$ such that the arguments of the factorials
are non-negative integers (zero is fine), also $m=j+k$,   and 
\beq
[\vec{l},\vec{j}]:&=&\left( \frac{(l_1-j)!}{(l_1+j)!}\frac{(l_2-k)!}{(l_2+k)!}(l+m)!(l-m)!\right)^{1/2}, \\ \nonumber
\Delta(\vec{l}):&=&\left( {(l_1+l_2-l)!(l+l_1-l_2)!(l+l_2-l_1)!(l_1+l_2+l+1)!}\right)^{1/2}.
\eeq

For purposes of this paper we need to specialize to $l_1=l_2=L$. The formula becomes:
\beq
C^{LL l}_{jk m} &=& (-1)^{L-j} \left( (l+m)!(l-m)! \frac{(2L-l)!(L-j)!(L-k)!}{l!l!(2L+l+1)!(L+j)!(L+k)!} 
\right)^{1/2} \\ \nonumber
&\times &\sum_r \frac{(-1)^r}{r!} \frac{(L+j+r)! (L+l-j-r)!}{(l-m-r)!(L-j-r)!(L-l+j+r)!}.
\eeq
The special case of this formula that we need for purposes of this paper is that
of one of the vectors being the highest weight. Thus, we choose $j=L$. It is easy to 
see that in this case the requirement that the argument of the factorial $(L-j-r)!$ is
non-negative truncates the sum over $r$ to the single term $r=0$. The
formula becomes:
\be
C^{LL l}_{Lk L+k} = \left( 
\frac{(2L)!(L-k)!(l+L+k)!}{(2L-l)!(2L+l+1)!(L+k)!(l-L-k)!} \right)^{1/2}.
\ee
Some special cases of this formula are:
\beq\label{cg-carlo}
C^{LL 2L}_{LL 2L} &=& \frac{1}{\sqrt{\rd_{2L}}}, \\ \label{cg-new}
C^{LL l}_{L-L 0} &=& \left( \frac{(2L)!}{(2L-l)!} \frac{(2L)!}{(2L+l+1)!} \right)^{1/2}.
\eeq
It is interesting to compute the asymptotics of the last coefficient in the regime $2L-l>>1$.
Using the Stirling formula one obtains:
\be 
C^{LL l}_{L-L 0} \approx \frac{1}{\sqrt{2L+l+1}}  \left( \frac{1-\frac{l}{2L}}{1+\frac{l}{2L}}\right)^{\frac{l}{2}}
\frac{1}{\left(1-\left(\frac{l}{2L}\right)^{2}\right)}^{L+\frac{1}{4}}.
\ee
Now, in the limit where $L\to \infty$, $l/2L \equiv x $ fixed and strictly smaller than one we have 
an exponential decay
\be 
C^{LL l}_{L-L 0} \approx \frac{1}{\sqrt{2L}}
\frac{1}{(1-x)^{\frac14}(1+x)^{\frac34}} \left(\frac{(1-x)^{x-1}}{(1+x)^{1+x}}\right)^{-L},  \ee
since one can easily check that the term inside the exponent is smaller than $1$ when $x<1$.
When $x$ is itself infinitesimal we have 
\be 
 \frac{(1-x)^{x-1}}{(1+x)^{1+x}}\approx e^{-x^{2}}.
 \ee

The general formula (\ref{cg-gen}) also gives the set of coefficients
relevant for the non-trivial Immirzi parameter case. These are given by:
\be
C^{j^{+} j^{-} k}_{j^{+} -j^{-}(j^{+}-j^{-})} = 
\left( \frac{(2j^{+})!}{(j^{+}+j^{-}-k)!} \frac{(2j^{-})!}{(j^{+}+j^{-}+k+1)!} \right)^{1/2}
\ee

\end{document}